\documentclass{article}

\def\be{ \begin{equation}}
\def\ee{ \end{equation}}
\def\ba{ \begin{eqnarray}}
\def\ea{ \end{eqnarray}}
  \def\F{{\cal F}}

  \newcommand{\ignore}[2]{#2}
\def\rellow#1#2{\mathrel{\mathop{\kern 0pt #1}\limits_{#2}}}
\def\o#1{\rellow{\circ}{#1}}
\def\circX{\rellow{\circ}{X}}
\def\circY{\rellow{\circ}{Y}}

\def\circinfty{\rellow{\circ}{\infty }}
\def\Mor{Mor}
\def\arrow{arrow}
\def\Arrow{Arrow}
\def\half{\frac 12}

\def\a{\alpha}
\def\b{\beta}

\def\pl{\partial}
\def\U{{\cal U }}

 \def\M{{\cal M}}

\def\A{A}
\def\E{{\cal E}}
\def\F{{\cal F}}
\def\O{{\Omega}}

\def\bfomega{\mbox{{\boldmath $\omega$}}}
\def\bfpi{\mbox{{\boldmath $\pi$}}}
\def\bfalpha{\mbox{{\boldmath $\alpha$}}}
\def\bfsigma{\mbox{{\boldmath $\sigma$}}}
\def\tildebfsigma{\mbox{{\boldmath $\tilde \sigma$}}}

\def\UC{\U (C)}
\def\bfU{{\bf U}}
\def\C{{\bf C}} 

\def\bitriangle{{\underline{\mathbf \Delta}}}

\newtheorem{theo}{Representation theorem}

\title{Pushing Einstein's Principles to the Extreme
\thanks{Carg\`ese lectures 1996, to appear in: G. 't Hooft et al (eds),
Quantum Fields and Quantum Space Time, Plenum Press 1997.
Work supported in part by the German Israel foundation}}
\author{Gerhard Mack\\
II. Institut f\"ur Theoretische Physik der Universit\"at Hamburg}

\begin{document}
\maketitle
 In these lectures I propose to push Einsteins principle of coordinate 
independence to the extreme in order to 
restrict the possible form of fundamental
equations of motion in physics. 
I start from nearly tautological system theoretic 
axioms. They provide a minimal
amount of {\em a priori} structure which is thought to be characteristic 
of human thinking in general.  
It is shown how formal discretizations of Maxwell and Yang Mills
theory in flat space 
and of general relativity in Ashtekar variables fit into this frame work.  
\section{What distinguishes truly fundamental physical theories?}
\label{fundamental}
The purpose of science is complexity reduction. We wish 
to understand a multitude of emergent phenomena starting from few basic 
principles. This tells us  what criterion could
be used to distinguish between more and less fundamental physical theories.
A theory will be the more fundamental the less  structure is 
assumed {\em a priori}. This is plausible, because what is assumed
 is not explained. Among the  {\em a priori} structure will be
all the axioms of the mathematical theories that are used to formulate the 
theory. Typically such axioms are relations between mathematical objects, and
there is no {\em a priori} reason why  the relations postulated in some
arbitrary mathematical theory should find their correspondence in nature.

What would be the minimum {\em a priori} structure which we need to assume in
order to build on it a theory of the world? Certainly the structure of 
human thinking will need to be included, because we cannot avoid using it
in building our theories.  Could such an assumption be enough by itself? 
We do not have the power to prove that it is, but it is very interesting to
examine the 
question of how far one can go,
 and it brings us into contact with the whole 
history of human thought. 

Philosophers might object that "structure of human thinking" must mean 
logic, and Immanuel Kant had proven in his famous "critique of pure reason" 
that a theory which describes  observed  phenomena in the world cannot be 
deduced from logic alone. 

However, studies of linguists \cite{bar:1} (and everyday experience)
 reveal that
 the structure of human thinking is not adequately represented by logic.

Moreover, Kant assumed too much structure {\em a priori},
including his version of the Aristotelian categories,
\footnote{In Hofstadters book "G\"odel, Escher, Bach"
 an intriguing interpretation of
the notion of enlightenment in Zen-Buddhism is proposed. Briefly it amounts 
to transcending below the level of Aristotelian categories to a more
 fundamental level of mental activity - thinking free of
(Aristotelian !) categories. In a way we are attempting something 
like that when we start from general systems and name things which would
 belong to different  categories. The Aristotelian categories are not 
{\em a priori} here, in contrast to Kant. In particular, the properties of 
{\em space} are not a priori}  
 and this leads to 
too many
possibilities of what can be thought of. Einsteins general relativity principle
amounts to postulating the {\em absence of a priori structure} as  we will see.
 It is well known that this relativity principle
 restricts  the form of the possible equations of motion very much. Here I try
to push this principle to the extreme. 
  
And last not least, 
the theory of complex systems has an entirely new selection principle,
inherited from quantum field theory and unknown to Kant, by which to select 
from all the theories which can be thought of those which give rise to 
{\em emergent phenomena} that could   in principle be observed.
I will come back to this later.

 One more objection must be 
answered. It might be argued that assumptions on the structure of human 
thinking are inappropriate as {\em a priori} assumptions, they should be 
deduced from neurophysiological data \cite{Edelmann}. But "deduced" is the 
wrong word. According to iron rules of logic, one is not allowed to use in 
a deduction that which is to be deduced. Building a theory of mental 
activity based on neurophysiological data makes use of the structure of
human thinking. One is trying to construct a self-consistent picture, and this 
can certainly produce very important insights,
 but it is not a deduction, and one
 {\em cannot} 
transcend to a level more fundamental than the structure of human thinking.

I will first state my basic assumption about the structure of human thinking 
informally. It will be made precise in the next section.\\[2mm]   
\noindent
{\bf Pre-Axiom:}{\em The human mind thinks about relations between things
 or agents.}
\footnote{There exists now a data base which categorizes and lists 
thousands of things the human mind thinks of, and the relations between them.
It was produced by Cycorp corporation with a view to commercial applications,
see the entries "The Cyc Technology" and "The Upper Cyc Ontology"
under http:///www.cyc.com , especially   http://www.cyc.com/cyc-2-1/toc.html 
(status Sept 96).     
}

Relations will be regarded as directed. The traditional notation in logic is
$aRb$ for a relation $R$ of $a$ to $b$. I prefer to use the notation 
which is now customary in mathematics (category theory), where one denotes
objects (things, agents) by capital letters X,Y,... and arrows (relations) by 
small letters $f,g,...$, and where 
$f:X\mapsto Y$ stands for a relation from $X$ to $Y$. 

It will be assumed as a defining
 property  of relations\footnote{It is said that string theory tries to 
construct geometry
from extended objects. But what means "extended" before there is space? 
It can only mean a property which is going to be interpreted as 
"being extended" after
space has been constructed. In the present frame work, {\em being a relation}
is such a property.}
 that they can be composed. 
If there is a relation $f:X\mapsto Y$ from $X$ to $Y$ and a relation
 $g: Y\mapsto Z$ from $Y$ to $Z$, then this defines a relation, denoted 
$g\circ f: X\mapsto Z$ from $X$ to $Z$. Think of a friend of a friend, or of 
a brother in law which is the husband of a sister. There can be relations from
$X$ to $X$; among them is the identity $\iota_X$ of $X$ with itself. 

Typically, a relation $f:X\mapsto Y$ from $X$ to $Y$ specifies 
 relation in the opposite direction, denoted $f^{\ast}:Y \mapsto X $. If 
$X$ is the wife of $Y$ then $Y$ is the husband of $X$. 

The objects of a system can themselves be systems, i.e have internal structure.
In this way, a general frame work for the discussion of self organization   
in complex systems is obtained. 

A generalized notion of {\em } locality will be built into the axioms.
 We know since the
discovery of Faraday's Nahewirkungsprinzip in the last century that 
fundamental physical laws relate only physical quantities at infinitesimally 
close points of space time. In discretized theories, the notion of
 infinitesimally close is replaced by nearest neighbor relations;
 this specifies a graph (e.g. a lattice) and singles out certain relations as
fundamental. 
All other relations can be composed from fundamental ones. The fundamental 
relations will be called {\em links}. 

This notion of locality leads to a definition of the notion of {\em emergence}
which is a key concept of complex systems theory. Emergence is the 
appearance of nonlocal phenomena as a consequence of local laws.
Propagation of electro-magnetic waves is an example, and also the 
reproduction fork dynamics which models the replication of DNA in cells (see
ref. \cite{mac:3} and figure \ref{reproFork} below).  

These basic assumptions will be subsumed in the axiomatic definition of a {\em 
system}; mathematically it is both a category and a graph. The relations are 
the arrows of the category, and there is a *-operation on arrows.

The objects are actually of secondary importance. They can be reconstructed 
when one knows which arrows can be composed, and what is the result of the
composition.

Classically, the state of the world at one time (and also the world
{\em sub specie aeternitatis}) is assumed to be described by a system of this
 kind. Quantum mechanically, there is a wave function which assigns a 
complex amplitude to systems.  

There are no numbers in this to begin with, and no 
arithmetic operations. One cannot make mistakes of $2\pi $ 
in the fundamental equations because there is  no $\pi$.
 The only substitute for arithmetic operations is the composition $\circ$ 
of relations. As a result, truly fundamental physical laws 
- those that can be stated in 
this language -  cannot contain any 
(dimension-less) free constants.
Also one cannot "add" physical theories (e.g. Einstein + Maxwell)
in a familiar way.

It will be seen later how one 
comes to correspondences with quantitative theories in the first place. 

 Coordinates are numerical encodings of positions in some space.    
The absence of {\em a priori} numerical structure is a way to push 
coordinate independence to the extreme.

How does one build a theory of the world on so little 
{\em a priori} structure? It proceeds in two steps

1. Name things,

2. Make statements about named things. 

\noindent In this paper I will be chiefly concerned with fundamental physics.
 The "things" which will be named and examined will be

electro-magnetic fields and Yang Mills fields,

space (in the sense of space-like hyper-surface of space time),

matter (Dirac fields).

To explain the naming step, it is necessary to distinguish between two
 different types of physical laws in the traditional formulation. 

First, there are laws which constrain the state of a part of the world at 
one time. 
Gauss' law in electrodynamics is a most important example of such a law. 
In a canonical formalism these laws are called {\em constraints}. All the
 fundamental physical theories, including general relativity,  
are gauge theories, and they all obey nontrivial constraints.  
There are further properties which can be read off the state at one time, and
which are preserved in time.
 I will count them among the constraints. 
It will be seen that our {\em a priori} structural assumptions, as poor as
 they are, provide for a gauge group (or a substitute for it) 
which can be read off the initial state, and for a notion
of gauge invariants which determines what could be observable in a particular
kind of system.  The named things in the above list will be systems which 
are distinguished by
 the validity of constraints which are characteristic for them.  The 
statement of the constraints must be meaningful, given only the {\em a priori}
 structure which is furnished by the axiomatic properties of a system. 
 
One may ask the philosophical question whether the constraints are
 really physical laws, or
just denominations. This brings us back to the discussion above. The
 principle of emergence may single out some of the possible constraints 
as physical laws because such properties of systems are the only ones which
can be observed at a  macroscopic level.

Secondly there are laws which govern the dynamics (time development)
 of a system. 
I will seek dynamical laws which are universal in the sense that they
 can be stated
in a meaningful way for any system whatever. This is a very restrictive 
requirement on a truly fundamental dynamical law, because there is so
little {\em a priori} structure which can be used to write down
 an equation of motion.

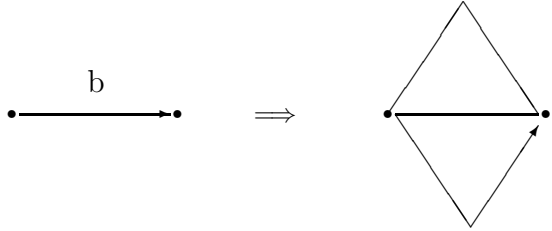
\begin{figure}
\begin{center}
\setlength{\unitlength}{1mm} %
\begin{picture}(200,25)(-25,36)
\put(10,53){{\large b}}
\put(1,50){\vector(1,0){20}}
\put(0,50){\circle*{1}}  
\put(22,50){\circle*{1}}  
\put(35,49){\makebox(0,0)[b]{\smash{$\Longrightarrow$}}}
\put(71,50){\circle*{1}}
\put(50,50){\circle*{1}}
\thinlines
\put(50,50){\line(2,3){10}}
\put(60,65){\line(2,-3){10}}
\put(70,50){\line(-1,0){19}}
\put(51,50){\line(2,-3){10}} 
\put(61,35){\vector(2,3){9}}
\end{picture}
\end{center}
\caption{The universal equation of motion of fundamental physics. The symbol
$\Longrightarrow$ symbolizes the effect of one time step. There is a product 
over all triangles which share the link $b^{\ast}$ ($=b$ with opposite 
orientation). A variant of the equation exists which has the orientation of 
the triangles reversed. The gauge covariant massless  Dirac equation is a 
special case; it governs the evolution of links $b$ to or from $\infty$. }
\label{fundamentalEqMotion}
\end{figure}

\begin{figure}
\begin{center}
\setlength{\unitlength}{1mm} %
\begin{picture}(200,30)(-25,22)
\thicklines
 \put(1,50){\vector(1,0){20}}
\thinlines
\put(0,50){\circle*{1}}   
\put(22,50){\circle*{1}}  
\put(35,49){\makebox(0,0)[b]{\smash{$\Longrightarrow$}}}
\put(71,50){\circle*{1}}
\put(50,50){\circle*{1}}
\thinlines
\put(52,51){\vector(1,0){17.25}}
\put(60,65){\line(2,-3){9.25}} 
\put(60,65){\line(-2,-3){10}}
\put(50,50){\line(1,0){20}}
\put(70,50){\line(-2,-3){10}}
\put(60,35){\line(-2,3){9.25}}
\put(51.25,49){\vector(1,0){17}}


\thinlines
\put(1,20){\vector(1,0){20}}
\put(0,20){\circle*{1}}  
\put(22,20){\circle*{1}}  
\put(35,19){\makebox(0,0)[b]{\smash{$\Longrightarrow$}}}
\thicklines
\put(70,20){\vector(-1,0){20}}
\thinlines
\put(50,22){\vector(1,0){20}}
\put(50,18){\vector(1,0){20}}
\put(50,20){\vector(0,-1){2}}
\put(70,22){\vector(0,-1){2}}
\put(48.6,20){\circle*{1}}  
\put(71.4,20){\circle*{1}}  
  
 \end{picture}
\end{center}
 \caption{Maxwell Equations of Electrodynamics. The Yang Mills equations
 of general gauge field theories have the same form. 
It involves a product over all triangles which share the horizontal link. 
In the presence of Dirac matter, the triangle can have a tip at $\infty$.}
\label{Maxwell}

\end{figure}
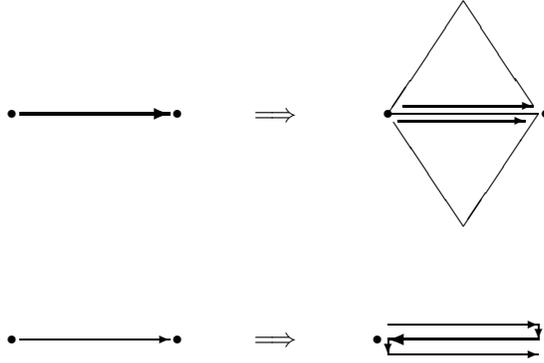

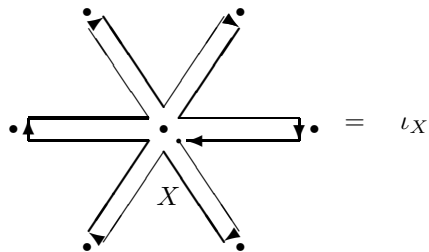
\begin{figure}
\begin{center}
\setlength{\unitlength}{1mm} %
\begin{picture}(60,30)(-24,-17)
\thicklines 
\put(-1,-10){$X$}
\put(24,0){= \ \ \ $\iota_X $}
\put(0,0){\circle*{1}}  
\put(2,-1.5){\circle{0.5}}  
\put(2,1.5){\line(2,3){8}}
\put(-2,1.5){\line(-1,0){16}} 
\put(18,-1.5){\vector(-1,0){15}} 
\put(-2,-1.5){\line(-2,-3){8}} 
\put(0,3){\line(-2,3){8}}
\put(0,-3){\line(2,-3){8}}  
\thinlines
\put(2,1.5){\line(1,0){16}}
\thicklines
\put(18,1.5){\vector(0,-1){3}}
\thinlines
\put(-2,1.5){\line(-2,3){8}}
\thicklines
\put(-10,13.5){\vector(3,2){2}} 
\thinlines
\put(2,-1.5){\line(2,-3){8}}  
\thicklines
\put(10,-13.5){\vector(-3,-2){2}} 
\thinlines
\put(-2,-1.5){\line(-1,0){16}} 
\thicklines
\put(-18,-1.5){\vector(0,1){3}} 
\thinlines
\put(0,3){\line(2,3){8}}
\thicklines
\put(8,15){\vector(3,-2){2}} 
\thinlines
\put(10.2,15.7){\circle*{1.0}} 
\put(10.2,-15.7){\circle*{1.0}} 
\put(-10.2,15.7){\circle*{1.0}} 
\put(-10.2,-15.7){\circle*{1.0}} 
\put(-20,0){\circle*{1.0}} 
\put(20,0){\circle*{1.0}} 
\thicklines
\put(8,15){\vector(3,-2){2}}
\thinlines
\put(0,-3){\line(-2,-3){8}}  

\thicklines
\put(-8,-15){\vector(-3,2){2}} 
\thinlines
\end{picture}
\end{center}
\caption{Gauss law for Electrodynamics, Yang Mills theory and
 General Relativity. In the presence of Dirac matter, one of the points is
at $\infty$.}\label{Gauss}
\end{figure}

In this paper I concentrate on the conceptual issues. The precise form of the 
equations is still open to experimentation. 
There is an essentially unique first order equation of motion, cp. 
figure \ref{fundamentalEqMotion}. To accommodate second order dynamics, I admit
two different kinds of links  - essentially coordinates and momenta 
(or velocities). They are represented by thin and fat lines in the figures.
But this weakens uniqueness; unfortunately 
there are now several equations which can be written down.  
Universal forms of the 
equations of motion of Maxwell,Yang Mills and of Einstein are shown
in figures \ref{Maxwell} and \ref{GRGmotion}. One sees that they fit on the
template, figure \ref{fundamentalEqMotion}, but they are not exactly the same. 
The constraints, figures \ref{Gauss},  \ref{GRGconstraints} are also not 
the same, but this is as it should be. 

It remains to be seen whether the two kinds of links can be fused into a 
single one
\footnote{In the Poincar\'e gauge theory approach to general relativity  one
tries to achieve such a fusion by interpreting vierbeins as vector potentials 
of the translation group. 
\cite{Hehl}}
 which satisfies one single universal equation of motion. 
 The universal law is supposed to specialize to the known fundamental
dynamical laws when applied to states which satisfy the appropriate
 constraints. 

There is no rigorous classification yet, and the investigation up to now are
 at the level of formal discretizations of known continuum theories. The
indications from the available evidence are that the following theories 
admit a universal formulation in the system theoretic frame work

1. General relativity with or without massless Dirac matter fields

2. Yang Mills theory in flat space with or without massless Dirac matter fields

But Einstein Maxwell theory (or Yang Mills theory in curved space) does not 
appear to fit; it is not unified enough. Also a cosmological constant,
 fundamental masses or fundamental Higgs fields do not fit.
The problem comes from explicit factors $\sqrt (-g)$ and $ g^{\mu \nu}$ which 
cannot be absorbed.
I will describe below a universal formulation of 
Maxwell- and Yang Mills equations, of
the Einstein equations, and of the massless Dirac equation. 

Discretizations of super-symmetric theories have not been investigated yet.  
They ought to be investigated because they may offer the best chance of
leading to emergent phenomena by virtue of cancelations of 
divergences at short distance which one encounters when one tries to enforce 
long range effects of short range interactions.   
%
%
%
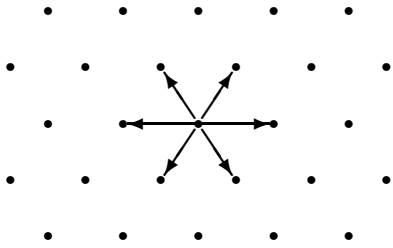
\begin{figure}
\begin{center}
\setlength{\unitlength}{0.5mm} %
\begin{picture}(130,45)(-17,10)
\thicklines
\multiput(10,20)(20,0){5}{\circle*{2}}
\multiput(10,80)(20,0){5}{\circle*{2}}
\multiput(0,65)(20,0){6}{\circle*{2}}
\multiput(0,35)(20,0){6}{\circle*{2}}
\multiput(10,50)(20,0){5}{\circle*{2}}
\put(51,51.5){\vector(2,3){8}}
\put(49,51.5){\vector(-2,3){8}}
\put(51,48.5){\vector(2,-3){8}}
\put(49,48.5){\vector(-2,-3){8}}
\put(49,50){\vector(-1,0){18}}
\put(51,50){\vector(1,0){18}}
\end{picture}
\end{center}
\caption{triangular lattice}\label{grid}
\end{figure}
%

 
\setlength{\unitlength}{0.8mm} 

\begin{figure}
\begin{center}  
\begin{picture}(200,25)(-70,-5)
\thinlines
 
\put(-12.5,-1){$\Rightarrow \prod$}   

 \thinlines
\put(-35,0){\vector(1,0){20}}

\thicklines
\put(0,3){\vector(2,3){10}}

\thinlines
\put(1.9,2.3){\vector(-3,2){2}}
\put(10,15){\line(-2,-3){8.6}}
\thinlines
\put(0,3){\line(3,-2){1.8}}

\put(10,15){\line(2,-3){10}}
\put(10,18){\line(2,-3){13.2}}

\put(3,0){\vector(1,0){17}}
\thinlines
\put(23.2,-1.7){\vector(-1,0){20.2}}
 
\put(25,-1.7){$\circ$}
\put(30,-1.7){\vector(1,0){17 }}
\end{picture}
\end{center}

\begin{center}  
\begin{picture}(200,25)(-95,-5)

\thicklines
\put(-60,-2){\vector(1,0){20}}
\thinlines
\put(-40,0) {\vector(-1,0){20}}
\put(-40,-2) {\vector(0,1){2}}
\put(-38,-2){$\Rightarrow$}

\thicklines
\put(-32,-2){\vector(1,0){20}}
\thinlines
\put(-12,0) {\vector(-1,0){20}}
\put(-12,-2) {\vector(0,1){2}}
\put(-10.5,-2){$\circ \prod$}


\thinlines
\put(21.5,-2){\vector(-1,0){22.7}} 
\thicklines
\put(10,15){\vector(2,-3){11.3}} 
\thinlines
\put(0,0){\vector(2,3){10}} 
\thicklines
\put(10,18){\vector(-2,-3){10}} 

\put(20,3){\vector(-2,3){10}} 
\thinlines 
\put(10,21){\vector(+2,-3){10}} 
\thicklines
\put(-2,3){\vector(2,3){12}} 
\put(20,6){\vector(0,-1){3}} 

\put(0,3){\vector(0,-1){3}} 

\end{picture}
\caption{Equations of motion of general relativity.
The product is over all triangles which share the horizontal link.}
\label{GRGmotion}
\end{center}
\end{figure}
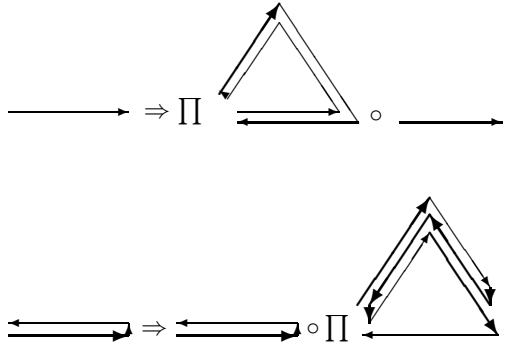

\setlength{\unitlength}{1mm} 
%
%
\begin{figure}
\begin{center}
\begin{picture}(200,25)(-55,-5)
\thinlines
\put(-15,0){ $\prod $ tr}

\put(10,15){\line(2,-3){10}} 
\put(20,0){\line(-1,0){20}}
\thicklines
\put(0,0){\vector(2,3){10}}
\put(10,2){{\large i}}

\put(25,0) {$=1$}
 \end{picture}
 
\begin{picture}(200,25)(-55,-5)
\put(-15,0){$\prod $ tr}

\put(10,18){\line(-2,-3){10}}
\thicklines
\put(10,15){\vector(-2,-3){8.6}}
\thinlines
\put(0,3){\line(3,-2){1.8}}

\put(10,15){\line(2,-3){10}}
\put(10,18){\line(2,-3){13.2}}

\thicklines 
\put(3,0){\vector(1,0){17}}
\thinlines
\put(3,-1.7){\line(1,0){20.2}}
\put(3,0){\line(0,-1){1.7}}

\put(25,0){$= 1$}

\put(-0.85,-0.85){\circle*{0.5}}
\put(-0.35,-3.0){{\large x}}
\end{picture}
\caption{vector and scalar constraint of general relativity. There is a 
product over triangles which share the link $i$ and the corner $x$, 
respectively. In $2+1$ dimensions there is a simpler version, cp. 
figure \ref{GRGconstraint3d}.}\label{GRGconstraints}
\end{center}
%
%
%
\begin{center}
\begin{picture}(200,25)(-55,-5)
\thinlines

\put(0,0){\circle*{1}}
\put(10,15){\line(2,-3){10}} 
\put(20,0){\vector(-1,0){18.5}}
\put(0,0){\line(2,3){10}}
\put(-2,-4){$X$}

\put(25,0) {$=\iota_X$}
 \end{picture}
\end{center}
\caption{Simplified form of vector + scalar constraint for $2+1$-dimensional 
gravity. The equality must hold true for every thin-lined triangle}
\label{GRGconstraint3d}
\end{figure}
\subsection{Einstein's principles}
\label{EinsteinPrinciples}
Let me pause to discuss how Einstein's principles fit in with the 
philosophy.
 
The two underlying principles of Einstein's General Relativity are the 
principle of
relativity, or general covariance, and the equivalence principle. 
When appropriately interpreted these principles are also operative in the 
gauge theories of elementary particle physics (modulo troubles with the Higgs 
sector)\cite{mac:2}. 
It is well known how these principles constrain equations of motion. 

The principle of relativity is a statement of {\em absence of a priori 
structure.} 
Before general relativity it was thought that  space has  an 
{\em a priori} structure which defines the notion of a straight line.
This is equivalent  to 
an {\em a priori} defined possibility of
comparing directions at different points in space. This {\em a priori} 
structure is abandoned in general relativity
and in gauge theory.  To compare tangent vectors
(or vectors in color space)  at different 
points of space time one must use parallel transport of vectors from
one point to the other, and the result depends both
on the path along which one transports, and on a connection 
($Sl(2,{\bf C})$-gauge field in relativity)
which is  {\em dynamically} determined as a solution of equations of motion. 
Gauge covariance follows.
 
In the traditional formulation of general relativity, the principle of 
relativity is not 
pushed to its logical conclusion, though. The assumption of an {\em a priori
structure} of space time as a differentiable manifold
 means that one assumes an 
a priori definition of straight line {\em in the infinitesimally small}. 
It has been 
suspected for a long time that this is an unreasonable assumption when it 
comes to 
physics at the Planck scale. In the traditional formulation, general 
covariance demands
 that there should be no preferred coordinate system. 
But one assumes an {\em a priori}
defined preferred {\em class} of coordinate systems, 
viz. smooth coordinates. General 
covariance is then interpreted to mean that the fundamental equations of the 
theory retain their form under transformations of coordinate systems 
{\em within the preferred class}. If one pushes the principle of coordinate 
independence to its logical conclusion, {\em the fundamental equations should 
make sense without any reference to coordinates whatever}. 

The principle of equivalence asserts that the motion of material bodies is 
free in a
local Lorentz frame. The notion of free motion makes essential reference to an 
{\em a priori} defined notion of straight line in the infinitesimally small. 
But when matter is described quantum mechanically, the notion of
"straight ahead" in the infinitesimally small is no longer needed. Newton's 
law gets replaced by a Schr\"odinger equation which involves a gauge invariant
Laplace or Dirac operator. To define it one needs only the appropriate 
parallel transporters, plus linearity which is supplied by the principles of 
quantum mechanics. 
In conclusion it is reasonable to hope that a sufficiently strong principle of 
coordinate independence alone should be sufficient to single out the truly 
fundamental dynamical laws in physics. 

Along the way, principal fiber bundles will go away. 
Mathematical physicists tend to think that principal fiber 
bundles are the essence of gauge theory, but it is not so. 
The definition of a global
multiplication from the right with elements of the structure group is 
is an a priori {\em global} structure. It amounts to postulating certain 
invariants. It is contrary to the spirit of gauge theory 
 which emphasizes locality. The principle fiber bundle structure provides for 
an {\em a priori} tensor product of representations
 which commutes with parallel transport. But  different tensor products
 are used in gauge theories with quantum gauge groups \cite{alex:1}; they are
 used in some models of quantum space time \cite{schom:1}.

\section{System theoretic foundations}
\label{system}
Motivated by the pre-axiom of the previous section I will now give a formal
 definition of a system. Following the terminology in category theory, the
agents of a system will be called {\em objects} and the directed relations 
between them are called {\em arrows}. 

A {\em complete system} consists of {\em objects } $X$ and {\em  arrows}
  $f:X\mapsto Y$. $X$ is called the {\em source} and $Y$ the {\em target} of 
the arrow $f$. 
 Some of the arrows  are declared
fundamental and are called {\em links}.
They are typically denoted by $b$.  The following conditions are imposed.
\begin{enumerate} 
\item 
Arrows can be composed. With $f:X\mapsto Y$ and $g:Y\mapsto Z$, the arrow
$g\circ f: X\mapsto Z$ is defined. Composition $\circ $ is associative.
\item 
To every object $X$ there exists a fundamental arrow $\iota_X:X\mapsto X$, 
called 
the identity arrow; $\iota_Y\circ f = f = f\circ\iota_X $ for every 
arrow  $f:X\mapsto Y$.
\item 
All arrows can be obtained by composing fundamental ones
\be f=b_n\circ ... \circ b_2\circ b_1 . \label{bbb}\ee
\item
To every arrow $f:X\mapsto Y$ there is an adjoint arrow $f^{\ast}:Y\mapsto X$ 
such that $f^{\ast\ast} = f$ and
$ (g\circ f)^{\ast} = f^{\ast}\circ g^{\ast} \ \ ,\
 \  \iota_X^{\ast} = \iota_X. $
\item
The graph whose vertices are the objects and whose links are the fundamental
 arrows is connected.  Equivalently: For every pair $X, Y$ there is an arrow
 $X\mapsto Y$
\end{enumerate} 
Mathematically, a complete system is both a category  and a directed graph.
The first two axioms are those of a category, the fourth one asserts 
the existence of adjoints of arrows. To specify the graph, certain 
arrows are singled out as links. 
The objects are the vertices of the graph and the 
fundamental arrows are the links of the graph. To every link, there is a link
in the opposite direction, but this requirement will be abandoned. 
 
In a {\em system}, the 4-th axiom is relaxed. I admit that this is motivated 
by hindsight.  
It is necessary to accommodate certain dynamical processes which are
very important in biology, such as DNA-replication \cite{mac:3}. 
The adjoints of some of the fundamental arrows $b$ 
are allowed to be  absent.
Thus, a {\em system} can be thought as being  obtained from a {\em complete
system} by declaring certain links as absent, and with them all arrows 
which can no longer be composed from fundamental arrows. However,
the absent adjoints can be added again in a unique fashion. 

We write $b^{\ast}=0$ if $b$ has no adjoint. There are  now 
two possibilities
\be  b^{\ast\ast} = b \ \ \mbox{or} \ \ b^{\ast\ast}=0.    \label{astast}\ee

Given a category, one needs to single out links to get a graph and a system. 
Let us examine the converse question: Given a directed graph,
 to what extent does it specify the system?

We can define a path from $X=X_0$ to $Y=X_n$ to consist of a sequence of links 
$b_0, b_1,... ,b_n$ such that the target of $b_i:X_i\mapsto X_{i+1}$ is 
also the source of $b_{i+1}$. 
Paths can be composed by juxtaposition. Adding the identity links, we obtain
 in this way a
category and thereby a system ${\bar S(G)}$ in a canonical way from a graph 
$G$. Therefore the assumption that relations can be composed is in fact a 
tautological one. It is useful because it institutionalizes the possibility 
that different paths may represent indistinguishable relations, as follows.

Suppose we start from a system $S$, and we reconstruct from its graph $G$
the system $ \bar S(G)$. It needs not be equal to $S$ because different 
paths may define the same arrow in $S$. The arrows in $S$ are in general 
equivalence classes of paths. Therefore, given a graph G, a system can be
 defined by specifying a {\em generating set of relations between links}. 
In the spirit of our locality principle, local relations are particularly 
important.

Two most important examples of such relations are as follows
 \ba 
 b^{\ast}\circ b = \iota_X &, & b\circ b^{\ast}= \iota_Y,  \label{rel}\\
 b_3\circ b_2\circ b_1 &=& \iota_X \label{rel1}
\ea
for all links $b:X\mapsto Y$ and for all 
triangles (i.e. loops of three links) from $X$ to $X$,
respectively. Interesting generalizations  will be encountered when we come to the Dirac equation. They differ only in some -signs. 

A further interesting type of relation is 
\be
\epsilon \circ b^{\ast } = b^{-1}\circ \epsilon 
\ee
where $\epsilon:X\mapsto X$ are {\em square roots of $-$ signs}.
 This could be used to characterize $sl(2,{\bf C})$ connections as appear
 in general relativity.

In the system theoretic frame work, a $-${\em sign} is a collection of links,
denoted $-\iota_X:X\mapsto X$, such that $-\iota_X\not= \iota_X$, but 
$(-\iota_X)\circ (-\iota_X)=\iota_X $, and $(-\iota_Y)\circ b 
= b\circ (-\iota_X)$
for all links $b:X\mapsto Y   $.

A system will be said to be {\em unfrustrated } if there is at most one 
arrow from $X$ to $Y$, whatever $X,Y$. Curvature in general relativity and 
field strength in gauge field theory are instances of frustration.

 We will also need a notion of isomorphism of systems because we 
will not distinguish between isomorphic systems. 

A {\em functor} $\F$ is 
defined as in category theory.
 It is a map from one system to another
 one which 
preserves identity and composition law.  
If $f:X \mapsto Y$ then $\F (f):\F (X)\mapsto \F (Y)$, $\F (\iota_X)=
 \iota_{\F (X)}$,
and $\F (g\circ f)= \F (g)\circ \F (f)$. It is not required that $\F$ maps
fundamental arrows into fundamental arrows, but 
it is postulated that $\F (f^{\ast})= \F (f)^{\ast} $.

Such a functor is called an {\em isomorphism}
of the system  if it has an inverse functor, and
if it maps fundamental arrows into fundamental arrows. 

\subsection{The language of thought}
Our assumptions on the structure of human thinking amounts to the postulate 
that the human mind manipulates objects and relations of systems by 
operations which are well defined as a consequence of the axiomatic 
properties of a system. If one entertains the notion that thinking uses some
sort of language, then one would be lead to calling the system theoretic 
frame work the {\em language of thought}. However, it is different from 
natural languages and from artificial languages including formal systems  in
one crucial aspect. All true languages have a serial structure. They are 
modeled on verbal utterances which are one word after another. General systems
have no serial structure.

Two questions arise naturally. How do properties of systems which occur in 
our mind during mental activity get translated into statements of a 
natural language, and where in the brain does the translation take place?

Related questions were raised by Raichle  \cite{Raichle}
 in his interpretation of recent 
neurophysiological experiments which localize types of mental activity in
the brain by a differential measurement of blood flow using PET (or Nuclear 
Magnetic Resonance).
\section{Gauge theory aspects}
\label{gauge}
I tried to make precise the idea that the human mind thinks about 
systems which consist of things and relations between them. 
It will presently be seen that 
this encapsulates the essence of gauge theories as we
know them in physics, in spite of the poverty of the assumed
{\em a priori} structure. 

Let $G_X$ consist of all arrows $ g:X \mapsto X$. They are called loops.
 Because of the composition law, $G_X$ is a semi-group. It will 
be called the holonomy semi-group or local gauge semi-group at $X$. 

A gauge transformation is a map of the system which takes every 
object $X$ into itself, and arrows $f:X\mapsto Y $ into new arrows 
$f^{\prime} :X\mapsto Y $ such that 
\be g(Y)f = f^{\prime } g(X) \label{gaugeS}\ee
for all arrows $f$ and a suitable choice of $g(Z)\in G_Z$ for all $Z$. 
Such a map is automatically functorial,
\footnote{In category theoretical language it is a functor which preserves
 objects  and which admits a natural transformation to
 the identity\cite{CWM}. }
i.e. 
$$   (g\circ f)^{\prime} = g^{\prime}\circ f^{\prime} $$

In  unfrustrated systems, the gauge semi-groups are trivial, i.e. they consist
 only of the identity $\iota_X$.

In our physical application, each arrow $b:X\mapsto Y$ will have an inverse 
$b^{-1}$ such that $b\circ b^{-1}= \iota_Y$ and $b^{-1}\circ b=\iota_X$.
 In this case
the local gauge semi-group is actually a group, and  it is independent 
of $X$ modulo isomorphism. It is called the {\em gauge group}.
 Gauge transformations
take the familiar form 
$$ f^{\prime} = g(Y) f g(X)^{-1} $$

\noindent Let us consider some examples of systems.\\[1mm]

{\bf Example 1} (triangulated manifold)
 {\em Consider a triangulated manifold. The objects of the system shall
 be the 0-simplices, and the links the 1-simplices. The adjoint link
is given by the 1-simplex with opposite orientation. Imposing the two relations
(\ref{rel}),(\ref{rel1}), the arrows $f:X\mapsto Y$ of the system will be 
the homotopy
classes of paths from $X$ to $Y$. The gauge group is the fundamental group of 
the manifold. The system is unfrustrated if and only if the manifold is
simply connected }

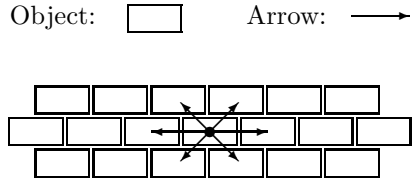
\begin{figure}
\begin{center}
\ignore{
\begin{picture}(5,6)(0,0)
}{
\unitlength1.0cm
\begin{picture}(6,1.5)(0,0.7)
\unitlength0.035cm
}
%
%
%
\put(10,80){\mbox{Object:}}
\put(55,77){\frame{\makebox(20,10){}}}
\put(100,80){\mbox{Arrow:}}
\put(140,83){\vector(1,0){22}}
%
%
\ignore{
\multiput(10,10)(22,0){7}{\frame{\makebox(20,10){}}}
}{}
\multiput(10,34)(22,0){7}{\frame{\makebox(20,10){}}}
\ignore{
\multiput(10,58)(22,0){7}{\frame{\makebox(20,10){}}}
}{}
\multiput(20,22)(22,0){6}{\frame{\makebox(20,10){}}}
\multiput(20,46)(22,0){6}{\frame{\makebox(20,10){}}}
 
 
%
%
\put(86,39){\circle*{4}}
\put(86,39){\vector(1,0){22}}
\put(86,39){\vector(-1,0){22}}
\put(86,39){\vector(1,1){11}}
\put(86,39){\vector(1,-1){11}}
\put(86,39){\vector(-1,1){11}}
\put(86,39){\vector(-1,-1){11}}
\end{picture}
 \end{center}
\caption{The structure of a brick wall}\label{fig:brickWall} 
\end{figure}

{\bf Example 2} (brick wall, see figure \ref{fig:brickWall}).
{\em The objects are the bricks, and the
 fundamental arrows are the translations which take  one brick to the 
position of a nearest neighbor. They can be composed to translations 
to other bricks  positions. The system is unfrustrated, and the gauge group
 is trivial.}

{\bf Example 3} (logical archetype) {\em 
The system has two objects, denoted $T$ and $F$ and three fundamental arrows 
other than the identities:
$$ e:T\mapsto F, \ \ e^{\ast}: F\mapsto  T, \ \ o = o^{\ast}: F\mapsto F $$ 
subject to the relations
$$ e\circ e^{\ast } = \iota_F, \ e^{\ast }\circ  e  = \iota_T, \ \ 
o\circ o = o. $$
The gauge semi-groups for $T$ and $F$ are isomorphic to the two element
 semi-group  $\{ \iota_F, o\}\cong \{ 1,0 \}$ with the usual
 multiplication law.
}

\subsection{Representations}
I will introduce a general notion of representation. 

In group theory, a representation is not simply a 
homomorphism from one group to another. 
 It is required that the representation operators are linear operators in a 
Hilbert space. As a result, there is an a priori defined multiplication
for them which is consistent with the linear structure in the Hilbert space in
the sense that the distributive law holds.

Similarly, models in model theory 
\cite{cha:1} are also a kind of representation.  They
are structure preserving maps whose images are sets.  

And in everyday life, an oil painting has some a priori structure
in addition to representing the structure of what is painted - it consists of
paint of some chemical composition on canvas. 

Motivated by this, {\em  representations} of a system $S$ will be defined as 
functorial maps  into some 
given system or into an element of a class of systems which come
 equipped with some 
characteristic additional structure. It is required that fundamental arrows
 are taken into 
fundamental arrows. 

{\bf Example} (logical representations)
{\em
 Equipped with a binary product $|$, 
\be T|T=F, \ \  T|F = F \ \ F|F = T \label{logic} \ee
the logical archetype (example 3 above)
 appears as the image of  representations ${\cal F}$ of 
systems which come equipped with a product (binary composition) $|$ of objects.
The notion of a product is understood to demand that 
 there are links $A\leftarrow A|B \rightarrow B $. 
\footnote{This generalizes  a corresponding construction in category theory 
\cite{CWM}}.  
Objects are interpreted as propositions,   
 links $A\mapsto B$ (other than the identity) are
 interpreted to mean " A excludes B", 
and $|$ means "neither nor". If a link's adjoint is its inverse, 
it gets interpreted as negation.   A representation assigns a truth value
$T$ or $F$ 
to every object (proposition); the representation property ensures that
 the rules of 
logic are obeyed, provided it is required that the representation preserves
composition, $\F (A|B) = \F (A)|\F (B)$, and the special links 
 $b: A|A \mapsto A $ are unitary (i.e. they obey  $b^{\ast}= b^{-1})$) so that
they will be  interpreted as negation.

Such representations may exist or not, and they may be unique or not.
One may consider logical representations of arbitrary systems with 
a product (binary composition $|$) but  it is natural to require unitarity
 $b^{\ast}=b^{-1}$ of the special links
 $b: A|A\mapsto A$. One writes $\neg A = A|A $.
If there is a subsystem of the form $\neg (A|\neg A)  \mapsto B$ then $\neg B$
 is interpreted as an axiom, because in any
 representation  $ \neg (A|\neg A ) $ is true. }

A representation may fail to exist because the axioms are contradictory. 
If the representation is not unique, then the truth of some propositions cannot
be decided from the axioms. 

\subsection{Representation of a system as a communication network}
Next I will state a representation theorem which will show that 
in spite of the nearly tautological character of our assumptions, 
all the essential structure of lattice gauge theory (on irregular lattices) is
encapsulated in it, 
{\em except} for the linearity of the charge - or color spaces whose 
elements are subject to parallel transport. 
The arrows will become maps, but not  necessarily linear maps.

{\bf Representation theorem: }{\it
 Every (finite) system admits a faithful representation as a
 network as follows: There are spaces $\O_X$ associated
with objects $X$ and arrows act as maps \ $f : \O_X\mapsto \O_Y , $ \
with $\iota_X=id$.
\/}

The construction of the
    space  $\O_X$ uses the sets of all arrows to and
from $X$. Details
are  given in Appendix A.

For now let us talk of one time. Then 
                    the maps $f$ may be interpreted  as  channels 
of communication. Time development (and acts of communication) 
 is only considered later.
 
The sets $\O_X$ need not be linear spaces and 
the maps $f$ need not be linear.
{\it Apart from this, the setup is as in  lattice
gauge theory.}. The objects $X$ may be elements of an irregular lattice
 but irregular lattices were
considered before. 

{\bf Scholium:}(Lattice gauge theory) 
\cite{Kogut}
{\em
In the Hamiltonian formulation of lattice gauge theory, space is a discrete
 lattice like figure \ref{grid}. In the continuum, one has vector
 potentials $A=A_idx^i $ which take their values in the Lie algebra of the
 gauge group. From them, parallel transporters along paths $C $ in space
from $X$ to $Y$  are
constructed as path ordered products $u(C)=P \exp (-\int_C A)$.
They map the fiber $\Omega_X$ of a vector bundle at $X$ into the fiber
 $\Omega_Y$ at $Y$.  In lattice gauge theory, the parallel transporters
 along the links of the lattice are the basic variables of the theory.  
Finite difference versions of covariant derivatives are constructed with the
 help of these parallel transporters.

Values of matter fields $\Psi(X)$ could be interpreted as elements of
$\Omega_X$ but we will prefer to regard them as maps (i.e. links) from some
"flavor space" $\Omega_{\infty}$ to $\Omega_X$ in the later
 discussion of the Dirac equation. 

The equations of motion and Gauss' law 
are the same as in the continuum, except that 
finite difference covariant derivatives in space are to be used. 
One may go on to discretize also time, which means that also time derivatives
get discretized.}

Let me emphasize  that the representation theorem constructs a space 
$\Omega_X$ for every object $X$ {\em but it does not attribute a state 
$\xi \in \Omega_X$ to the 
objects}. The objects  have no state. Dynamics
consists of structural transformations,
not of changes of states of objects. 
 This is a big difference to cellular automata.
Nevertheless there is a connection. At the level of {\em effective theories} 
which operate on larger scales, the objects can be  systems themselves, and so
they have internal structure. Changes of this internal structure could be 
interpreted as changes of a state of the object. 

It is appropriate to cite also the computer pioneer Konrad Zuse's work on 
"computing space" \cite{Zuse} for similarity in spirit. The frame work which 
I use here to discuss fundamental physics is also
 employed as a tool in  massively parallel computing \cite{mac:3}.  

Let me clarify that I do not regard 
values of matter fields $\Psi(X)$ in gauge field theories as elements of
 $\Omega_X$ because 
otherwise it would be impossible   to find a universal equation of motion
 for the links in which the values of
 the matter fields would enter. This brings us to the next topic.

\section{Universal dynamics} 
\label{universal}
Next
we turn to the time development $t\mapsto S_t$ of a system.
 Sorin Solomon proposed to
call it "drama". It is supposed to be governed by an equation of motion.

According to our guiding principle, the most fundamental equations of 
motion should have the property that they can be formulated purely 
within the frame work provided by the language of thought, without need 
for any further a priori structure. In other words, they should be 
meaningful for every system whatever. 
I will call this a universal dynamics. 

Another consideration  leads also to universal dynamics:
A state should contain all
necessary information about its time development in itself, without
need for further extrinsic specification. Different kinds of systems should
be distinguished by different properties of the initial states. 

Gauge invariance is automatic in a universal dynamics because there is no 
intrinsic way to distinguish between isomorphic representations of a system.

I will consider dynamics in discrete time.
Dynamics in continuous time would require some assumptions of {\em a priori}
structure such as spaces $\Omega_X$ which are manifolds. 

 I will assume at first that
the dynamics is of first order, so that the system 
$S_t$ at time $t$ determines
the system $S_{t+1}$ at time $t+1$. Generalization to second order dynamics 
 will be considered later and reduced to the first order case, but with two 
kinds of links. 

In the spirit of the discussion of locality
 in section \ref{system} it is demanded 
that the dynamics is local in the following sense

Every object is descendent of some object $X$  and every link is descendent of
some  arrow
$f$ of the system one time step ago. Descendents
of $X$ are determined by $X$ and by the fundamental arrows of $X$ alone. 
Descendent links of $f$ are determined by $f$
if $f$ is fundamental, by source $X$ and target $Y$ of $f$,
 and possibly by the fundamental arrows to and from $X$ or $Y$. 
 
The formulation of a dynamics is a rule how a new system is to be made out of 
a given one. It is supposed to have the stated locality properties. The 
possibilities of formulating such rules within the language of thought 
are very restricted. In fact, the innocent looking assumption of a system with
a {\em finite} number of links has introduced a priori structure of
countability. I exorcize it again
by not 
admitting the possibility of counting the number of links to an object. 
It should make no difference if several simultaneous links from
$X$ to $Y$ are regarded as one link. 

Basically there are three kinds of change with time, apart from death.
\begin{enumerate}
\item
Growth
\item
Motion
\item
change of composition law
\end{enumerate}
This classification applies not only to material bodies in space,
but in this paper we are only concerned with physics. 

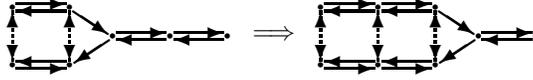
\begin{figure}
\begin{center}
\setlength{\unitlength}{0.001875in}%
\begin{picture}(1474,180)(123,460)
\thicklines
\multiput(1150,620)(0.00000,-16.47059){9}{\line( 0,-1){  8.235}}
\put(1150,480){\vector( 0,-1){0}}
\put(1150,620){\vector( 0, 1){0}}
\put(1310,470){\circle*{14}}
\put(1310,630){\circle*{14}}
\put(1160,640){\vector( 1, 0){140}}
\put(1000,460){\vector( 1, 0){140}}
\put(1150,470){\circle*{14}}
\put(1150,630){\circle*{14}}
\put(1000,640){\vector( 1, 0){140}}
\put(1140,620){\vector(-1, 0){140}}
\put(1140,480){\vector(-1, 0){140}}
\put(1300,620){\vector(-1, 0){140}}
\multiput(990,620)(0.00000,-16.47059){9}{\line( 0,-1){  8.235}}
\put(990,480){\vector( 0,-1){0}}
\put(990,620){\vector( 0, 1){0}}
\put(1580,540){\vector(-1, 0){140}}
\put(1440,560){\vector( 1, 0){140}}
\put(860,540){\makebox(0,0)[b]{\smash{$\Longrightarrow$}}}
\put(990,470){\circle*{14}}
\put(1300,480){\vector(-1, 0){140}}
\put(1160,460){\vector( 1, 0){140}}
\multiput(1310,620)(0.00000,-16.47059){9}{\line( 0,-1){  8.235}}
\put(1310,480){\vector( 0,-1){0}}
\put(1310,620){\vector( 0, 1){0}}
\put(990,630){\circle*{14}}
\put(1590,550){\circle*{14}}
\put(1420,540){\vector(-3,-2){ 96.923}}
\put(130,470){\circle*{14}}
\put(140,640){\vector( 1, 0){140}}
\put(280,620){\vector(-1, 0){140}}
\put(300,620){\vector(3, -2){ 96.923}}
\put(410,550){\circle*{14}}
\put(130,630){\circle*{14}}
\put(290,470){\circle*{14}}
\put(290,630){\circle*{14}}
\put(570,550){\circle*{14}}
\put(730,550){\circle*{14}}
\put(400,540){\vector(-3,-2){ 96.923}}
\put(560,540){\vector(-1, 0){140}}
\put(580,560){\vector( 1, 0){140}}
\put(1430,550){\circle*{14}}
\put(1320,620){\vector(3, -2){ 96.923}}
\put(720,540){\vector(-1, 0){140}}
\put(280,480){\vector(-1, 0){140}}
\put(140,460){\vector( 1, 0){140}}
\multiput(130,620)(0.00000,-16.47059){9}{\line( 0,-1){  8.235}}
\put(130,480){\vector( 0,-1){0}}
\put(130,620){\vector( 0, 1){0}}
\put(420,560){\vector( 1, 0){140}}
\multiput(290,620)(0.00000,-16.47059){9}{\line( 0,-1){  8.235}}
\put(290,480){\vector( 0,-1){0}}
\put(290,620){\vector( 0, 1){0}}
\end{picture}
\end{center}
 \caption{Reproduction fork dynamics - a universal copy machine for systems.
 A pair of links without adjoints to and
 from an object $X$ is called a fork. The
 presence of a fork causes $X$ to be copied. The
 bidirectional links get split to become forks and the
 two halves are divided among the copies
 of $X$. The links which had no adjoint before get one. 
Once a copy process is started at some initial object
 $X_0$, the forks travel through the whole system and
 one gets two copies of the system as a result.  (The
 dotted arrows are only there to
indicate the fact that the objects are copies of each
 other.) This works for systems of completely arbitrary topology. 
}\label{reproFork}

 \end{figure}

I speak of growth if  there is copying of objects or of links, or if 
adjoints $b^{\ast}$ of links $b$ are newly produced. 
There can also be fusion of isomorphic subsystems under some conditions. 
The aforementioned
reproduction fork dynamics   is a universal dynamics. It models DNA
 replication \cite{alb:1} but has also much more general copying capabilities. 
It is a local dynamics which propagates a copy-process in such a way that 
systems of completely arbitrary topology can be copied. It is shown 
in figure \ref{reproFork}. For further explanation see ref. \cite{mac:3}. 
Locality is important because enzymes in a biological system act locally. 

In principle there exists the possibility of a change in the composition law
$f\circ g \Rightarrow f \circ s\circ g $, where $s$ is a loop. 
But I will not enter into a discussion of this possibility here.

Here I will be chiefly interested in motion.
 It consists of  changes 
of arrows. 
\footnote{As I mentioned before, the 
 objects are secondary. This is so because they can be recovered
  from the arrows and
the composition law according to the representation theorem. This is true
up to isomorphism. Isomorphic systems are not regarded as different. There
is no intrinsic way of distinguishing between them. }
This includes changes of relations of an object $X$ to itself; these
relations could be regarded as properties of $X$.

The possibilities are very limited.
 How can a link $b^{\prime}:X\mapsto Y$ of the 
system at time  $t+1$ arise? It can only have been composed from links of
the system $S_t$ at time $t$.
 (Creating new links e.g. by taking adjoints  would be regarded as growth).
But all that can be made by composing links is an arrow of the category.
So the rule has the form
\be b^{\prime }= f\label{genMotion}\ee
where $f$ is a possibly composite arrow of $S_t$. But this means 
that {\em the category does not change at all}.
\footnote{This reminds of Parmenides, the Greek father of ontology.
 He held that nothing can appear or
 disappear in the world because this would contradict the nonexistence of the 
nonexistent. Such changes are only apparent ones to man.} 
 The only change is in the specification which
arrows are considered as fundamental.
{\em Motion means that composite relations are declared fundamental.}
One can think of it as composition of links or bonds by objects
 which act as catalysts
in a manner which is familiar from chemistry, see figure \ref{enzymePic}. 
\setlength{\unitlength}{1mm} 

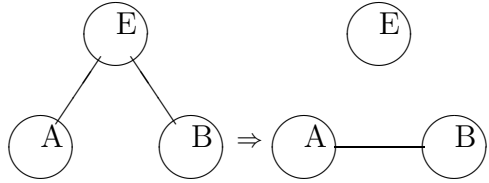
\begin{figure}
\begin{center}  
\begin{picture}(200,25)(-30,-5)

\put(10,15){\circle{8}}
\put(10,15){\large E }

\put(0,0) {\circle{8}}  
\put(0,0){\large A }
\put(2,3){\line(2,3){6}}
\put(20,0){\circle{8}}
\put(20,0){\large B }
\put(18,3){\line(-2,3){6}}

\put(26,0){$\Rightarrow$}

\put(45,15){\circle{8}}
\put(45,15){\large E }

\put(35,0) {\circle{8}}  
\put(35,0){\large A }
\put(39,0){\line(1,0){12}}
\put(55,0){\circle{8}}
\put(55,0){\large B }

\end{picture}
\caption{Catalysis in chemistry.
 Enzyme E binds molecules A and B. 
First a substrate-enzyme complex is formed where A and B are bound to E.
 Then the composite link between A and B 
is transformed into a fundamental one}\label{enzymePic}
\end{center}
\end{figure}
%

%
\begin{figure}
\begin{center}
\setlength{\unitlength}{0.001875in}%
\begin{picture}(1234,137)(243,540)
\thicklines
\put(1460,540){\vector(-1, 0){140}}
\put(480,660){\vector(-2,-3){ 60}}
\put(560,540){\vector(-1, 0){140}}
\put(420,560){\vector( 1, 0){140}}
\put(400,540){\vector(-1, 0){140}}
\put(1320,560){\vector( 1, 0){140}}
\put(860,540){\makebox(0,0)[b]{\smash{$\Longrightarrow$}}}
\put(1240,660){\vector( 2,-3){ 60}}
\multiput(1160,560)(16.47059,0.00000){9}{\line( 1, 0){  8.235}}
\put(1300,560){\vector( 1, 0){0}}
\put(1300,540){\vector(-1, 0){140}}
\put(1140,540){\vector(-1, 0){140}}
\put(1000,560){\vector( 1, 0){140}}
\put(260,560){\vector( 1, 0){140}}
\put(410,550){\circle*{14}}
\put(250,550){\circle*{14}}
\put(490,670){\circle*{14}}
\put(730,550){\circle*{14}}
\put(570,550){\circle*{14}}
\put(1310,550){\circle*{14}}
\put(580,560){\vector( 1, 0){140}}
\put(720,540){\vector(-1, 0){140}}
\put(1150,550){\circle*{14}}
\put(990,550){\circle*{14}}
\put(1230,670){\circle*{14}}
\put(1470,550){\circle*{14}}
\end{picture}
\end{center}
\caption{Interpretation of motion as transformation
 of a composite relation into a fundamental one. The
 objects which are connected by bidirectional links are
 interpreted as space points, and the other object as a
 particle (or as ``the idea of matter''). The link from 
the particle to a space point $x$ represents the
 relation of ``being at $x$''. Motion takes place
 when a composite arrow made from the  relation $b$ of
 the particle to its former position, and a relation of
 this space point to a neighbor is declared fundamental, while
 $b$ loses this status. $b$ remains in the category
 as a composite arrow.}\label{motion}

\end{figure}
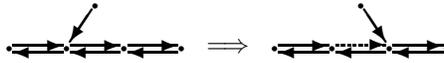

Such catalysis of relations  also plays a basic role in Spinoza's famous
 treatise  on ethics, {\em Ethica, ordine geometrico demonstrata} 
\cite{Spinoza}.

The interpretation of the motion of a point particle in space 
is shown schematically in figure \ref{motion}.

Let us return to  the analysis of the possibilities for $f$ in the 
general formula (\ref{genMotion}) for motion. 

The links in $f$ are restricted by the locality demand. The simplest 
possibility is as follows.
\be b\mapsto b\circ b^{\ast }\circ b . \ee 
This reminds of Hegel's dialectic process. 
Let us follow Hegel in speaking of a ``Denkbestimmung'' in place of a link. 
Then the process gets verbally described as follows.  
A Denkbestimmung (thesis) combines with its opposite (antithesis) to form a 
new,''more advanced'' Denkbestimmung (synthesis). Actually something
new is obtained only if $b^{\ast }$ is not the inverse of $b$. 

The dialectic process is truly a universal dynamics in the sense of the above
definition. But it is not the only one.
 Neither Maxwell's 
equations nor Einstein's are of this form. 

Other possibilities are found by making use of
 links $b_i$ to or from $X$ or $Y$
 other than the original link $b:X\mapsto Y $ and its adjoint $b^{\ast}$.
 Assuming, contrary to Hegel,
that  $b_i^{\ast}\circ b_i$ constructions
 yield nothing nontrivial, the useful 
links can only occur in triangles $\triangle_i$
which contain $b$ or $b^{\ast}$. The triangular paths 
 have the form  
\be \triangle_i =b^{\ast }\circ b_i^1\circ b_i^2 . \label{triangle}\ee
or adjoint of that. This involves links  $b^1_i: X\mapsto Z$, 
 $b^2_i: Z\mapsto Y$, $(Z\not= X,Y)$.
 Different triangles cannot be distinguished in an intrinsic way.
Therefore they will have to appear symmetrically in the rule.
Also we have no way of adding 
contributions. The only possible composition is with  $\circ$ . This leaves
us with the possibilities 
\be
b \mapsto b \circ \triangle_1\circ ... \circ \triangle_n \label{unieqa} 
\ee
or
\be
b \mapsto \triangle_1^{\ast }\circ 
... \circ \triangle_n^{\ast}\circ b \label{unieqb} 
\ee
where $\triangle_1$... $\triangle_n $ are all the different triangular paths
of the
form (\ref{triangle}).

This is the universal equation of motion of fundamental physics, modulo some 
complications which we will discuss below.

 The path on the right hand side of
 eq. (\ref{unieqa}) starts with a factor $b \circ b^{\ast }$  which could be 
omitted. 

A schematic graphical representation of eq.(\ref{unieqa}) is in
 figure \ref{fundamentalEqMotion}. 
\subsection{Universal conservation law}
\label{conserv}
In the absence of growth processes, a universal equation of motion implies 
substitutions $b\mapsto f$ which replace links by arrows which existed before 
in the category. Therefore the category $Cat (S)$ does not change
 in the course of time. Quantities $Q$ which are determined by
 the isomorphism class
 of the category are conserved. In particular the gauge group is time
independent. 

If there are growth processes, new objects which are copies of old
 ones may appear, and also new links which are copies or adjoints of old ones. 
In this case the conservation laws are more subtle. 

\section{Maxwell's equations} 
\label{MaxwellSection}

In general, a system is identified as {\em space} by the validity of
 certain constraints. This will be discussed below when we come
 to general relativity. 
 In this section, I am not interested in this aspect and
 I will assume that we know already what is a discretized flat
 space. Let us think of a triangular lattice like figure \ref{grid}. 

Maxwell's equations come in two groups. The first group states restrictions on
the initial state. The equation $div\, B = 0$ is automatically satisfied
through the introduction of a vector potential, and there
 remains only Gauss law,  $div \, E = \rho$. Gauss law in our
 language is shown in figure \ref{Gauss}

 The second group contains equations of motion. Their universal version
 is shown in figure (\ref{Maxwell}).

The Yang Mills equations have exactly the same form,
apart from an ordering problem which will be discussed. But the gauge group -
 which is a property of the initial state - is different. The
 equations retain their form in the presence of Dirac matter fields, but
 in this case one 
of the points in the diagram represents a flavor space (or point at infinity)
rather than a space point.  

I emphasize that these equations are generally meaningful, but they 
reduce to Maxwell's or Yang Mills equations only on a regular "flat" lattice.
 I do not know a universal version of the Maxwell or Yang Mills equations 
on curved space. 

The gauge group is part of the data which specify which particular aspect of
the world we are dealing with. In Electromagnetism, the gauge group
 is isomorphic to $U(1)$. This group admits a natural parameterization by real 
numbers $0... 2\pi$. The magnetic field is a an element of the gauge group
(parallel transporter around a loop) and so it acquires a numerical status. 
 Similarly the electric field  is a loop which involves parallel
transport forth and back at two successive times. 
In this way they both become  "quantities" in spite of the fact that
 the general frame work
knows no numbers. They are gauge invariant and therefore observable in
 principle. 

If the gauge group is noncommutative, there arises an ordering problem. 
In what order shall the triangles be traversed? The only reasonable
 answer is "at random".
This introduces some stochasticity which may be thought to be a remainder from
the quantum theory. 
In the formal continuum limit its effect disappears.
 
In the quantum theory,
the superposition principle furnishes a commutative operator $+$, and we
 can sum over triangles instead of composing triangles with the help of
 $\circ$. So the above ordering problem disappears, but in its place 
we have the "ordering problems" of quantum mechanics. A universal 
Schr\"odinger equation is described  in \cite{mac:1} and in
 section \ref{QSpaceTime} below. 

\section{General relativity}
Next I turn to the equations of motion and constraints of general relativity. 
The constraints are the properties which a system should have in order to be
interpreted as space (in the sense of space-like surface in space time).
Among the constraints is the selection of the gauge group 
 $Sl(2,{\bf C})$.  
In 2+1 dimensions it would be the covering group $Sl(2,{\bf R})$ of the 
3-dimensional Lorentz group instead. 

Also among the constraints is the existence of an invariant trace $tr$ 
which maps loops to numbers. 

The remaining constraints are shown in figure \ref{GRGconstraints}, and the
equations of motion in figure \ref{GRGmotion}.
They come from the 
 canonical formulation of general relativity in Ashtekar variables. 
For the readers convenience a brief review is given in Appendix B.

There exists also a version of the constraints which does not involve $tr$,
cp. \cite{AshtekarFootnote}. It is only equivalent under invertibility
 conditions on the dreibein. A similar reformulation exists in 2+1 dimension
where it takes the very simple form of a flatness requirement $F_{ij}(x)=0$.
\cite{Wael}
In our language this is eq.(\ref{rel1}) (figure \ref{GRGconstraint3d}) for
 thin links $b_i$. It says that thin 
loops around triangles equal the identity. 
\section{Interpretation of the equations of motion}
\subsection{Syntax and Semantics: Classification of initial states}
The graphical form of the Maxwell, Yang Mills,  and
Einstein-Ashtekar equations could have been written down by 
pre-Sokratic philosophers. They 
must surely have conceived of the idea that the world evolves by 
rearrangement of triangles. But they could not possibly have found the 
proper interpretation.

And some readers will no doubt be left in a state of perplexity by the claim 
that the universal equations of motion (\ref{Maxwell}) contain the Maxwell,
Yang Mills equation. The precise correspondence will be 
explained. But the perplexity itself is worth a comment. It makes it clear 
that the problem of describing nature and its laws is not solved yet by 
stating laws in the form of equations. These equations live on a purely 
spiritual level. They are syntactical rules. In addition one will ask for 
their meaning. This is the question of semantics. It asks how entities in
the equations correspond with phenomena which can be observed in nature. 
Physicists say the theory itself must determine what quantities are observable.
The question about  meaning   is therefore divided into two questions. 
The first questions is what quantities from the equations 
can be observed in principle. The second question is where to find these
quantities in nature. 

The first question has already been answered. Observations give answers 
yes or no to questions whether certain statements about a system are true. The
permissible statements are those which can be formulated in the language of
thought. It was pointed out that this implies that the observable 
quantities are gauge invariant. 

The second question is rarely considered in practice.  Typically equations 
in physics are written down in order to explain certain phenomena. In 
this case the observable quantities have already been fixed a priori by 
the scope of the investigation. 
In the present frame work the question is tied to the question how systems
specified as initial states identify themselves - e.g. as electro-magnetic 
fields to which Maxwell's equations of motion would apply, or as  space
(=space-like hyper-plane in space time) to which the equations of motion
of general relativity would apply. 
 The general answer is this. They identify themselves by properties which 
can be formulated in the language of thought, and which are 
preserved (in their totality) by the time development.
 Physicists call them {\em constraints.}
Semantics demands therefore that systems with these properties are given 
names in natural language. 

The question may arise whether all possible such properties will also occur in
 nature. The practical answer is that not all possible properties can be
expected to give rise to phenomena that can be observed {\em at scales  
which are very large} compared to scales set by
 the equations. Only properties which give rise to emergent behavior 
can be observed in practice. This shows that fundamental physics is part of
complex systems theory \cite{mac:4}
. 

In electrodynamics and Yang Mills theory, the familiar constraint is 
Gauss' law. It can be formulated in the present frame work in the form of
 statements ``loop $=$ identity''.  The result is shown in
schematic form in figure \ref{Gauss}.  

In addition there are further properties which assert the existence of 
invariants. The properties of this type are linearity of the state spaces 
$\Omega_X$ and of the maps $f$ between these state spaces (links $b$  =lattice
gauge fields), and the identification of the gauge group $G$ and of 
$\Omega_X$ as
a representation space of it. Linearity amounts to the existence of  
invariant operators for addition, 
$+:\Omega_X\times \Omega_X \mapsto \Omega_X$ and for
 multiplication with numbers
$ * : {\bf C}\times \Omega_X\mapsto \Omega_X$ with
associativity, commutativity and distributivity properties. [The definition of
the gauge group was described before.]
The defining property of invariants is that their
 parallel transport is path independent. 
As a result they can be globally defined in such a way that they commute
 with parallel transport. 

The gauge group in Electrodynamics is Abelian. This amounts to the statement 
$$ s\circ s' = s'\circ s  $$
for arbitrary loops $s,s':X\mapsto X$. 

The constraints in general relativity include the Gauss law. In addition there
are further constraints. Their statement involves an invariant $tr $ whose
 existence is also one of the constraints. It maps loops to numbers. 
The further constraints are shown in figures  \ref{GRGconstraints} 
The  equations of motion of general relativity are shown in figure \ref{GRGmotion}.

 All of this discussion is  at the level of formal discretizations of  the 
standard theories. 
\subsection{Interpretation of the universal form of Maxwell's equations} 
Maxwell's equations of motion are of second order once the vector potential 
is introduced. Therefore the initial state needs to specify both
 coordinates and velocities or momenta. As a result, there will be
 two kind of links. They 
will be printed thin and fat, 
 respectively. One could try to distinguish them by properties in
 an intrinsic way, e.g. by postulating that one kind has an
 adjoint and the other does not. 

In the continuum, the canonical variables are vector potential ${\bf A}(x)$ and
electric field ${\bf E(x)}$, and the Maxwell equations are 
\ba 
        \dot {\bf A} &=& -   {\bf E} \label{Adot}\\
         \dot {\bf E} &=& curl {\bf B}
\label{Edot}
\\ {\bf B}&=& curl {\bf A}.
\ea   
On the lattice one uses instead exponentiated 
quantities.  We may label the links  with target $x$ in some way 
by $i=\pm 1, \pm 2,...$.
  They have a direction and a length given by a vector $\hat i$. 
Choosing $X$ and $i$ will select a link $b_i(x)$. The opposite link to 
$b_i(x)$ is denoted $b_{-i}(x+\hat i)$. 
  Let $A_i(x)$ and $E^i(x)$ be
 the
components 
of ${\bf A}$ and ${\bf E}$ in the directions of the links  .

The thin links will be the parallel transporters
of lattice gauge theory. Assuming a vector potential which is
 smooth on the scale of the lattice spacing, the parallel transporters
 can be approximated,
\be U(b) = P\exp \left( - \int A_{\mu }dx^{\mu } \right) 
\approx  \exp \left(-  A_i(x+\hat i/2) \right)
\ \mbox{ for } b=b_i(X). \label{U}
\ee
The fat links are taken to be
\be
P(b) =  U(b)\exp \left(-\tau E^i(x)\right) \ .  
\ee
where $\tau $ is a discrete time step. 
$U(b)$ and $P(b)$ are complex numbers of modulus 1; the vector potential
and the electric and magnetic field are regarded as 
pure imaginary - otherwise a factor $i$ has to be put in the exponents. 
The $E$-variables on opposite links are related by
\be\exp \left(-\tau E^{(-i)}(x+\hat i)\right) U(b) =
  U(b)\exp \left( \tau E^i(x)\right) \label{adjP} \ee

These quantities will be functions of a discrete time $t$ . It follows from 
eqs.(\ref{Adot}) and (\ref{adjP}) that 
\ba
      U_{t+\tau}(b) &=& U_t(b)\exp\left( -\tau E^i(x)\right)\\
      &=& U_t(b)P_t(-b)U_t(b) \ , 
 \label{UplusTau}
\ea
which is the second of the universal Maxwell equations in 
figure \ref{Maxwell}. 
 
Consider now the triangles $\Delta$  in the first equation in figure
 \ref{Maxwell}.  In the limit of small lattice spacing, path ordering
  can be neglected and the 
parallel transporter around a triangle is given by the magnetic flux
$\Phi_{\Delta } $ through the triangle
\ba 
U(\Delta ) &=& P\exp (-\int_{\Delta} Adx ) = \exp ( -\Phi_{\Delta })
\label{flux}
\\
\Phi_{\Delta } &\approx & B^{\perp}\cdot (\mbox{area of } \Delta ) 
\label{magField}
\ea
where $B^{\perp}$ is the magnetic field perpendicular to the triangle. 
The two triangles $\Delta_1$ and $\Delta_2$ on opposite sides of the link
in figure \ref{Maxwell} have opposite orientation.  Therefore the factors 
$\exp \left( -\Phi_{\Delta_i} \right)$ will cancel, except for the 
effect of the change of the magnetic field component $B^{\perp} $ in the 
direction perpendicular to the link and to $B^{\perp}$.
 This change is part of the component of $curl B$ in the direction of 
the link -
all of it in 2 space dimensions. Taking the product over the pairs of 
triangles  in all directions perpendicular to the link, one gets
$\exp (-\gamma \hat i \cdot curl{\bf B_t}) $ where 
$\gamma $ is a dimensionful constant of geometric origin.  
The right hand side of the equation is this product multiplied with $U_t(b)$.
So the equation reads
\be 
P_{t+\tau} = U_t
\exp ( -\gamma \hat i \cdot curl {\bf B_t}) \ .
\ee
For suitable choice of the time step, $\tau=\sqrt\gamma$,
 this is an exponentiated form of the Maxwell equation (\ref{Edot}), 
because 
\ba P_{t+\tau}&=& U_{t+\tau}\exp\left( -\tau E^i_{t+\tau}\right) \\
&=& U_t \exp\left( \tau E^i_t\right) \exp\left( -\tau E^i_{t+\tau}\right)\\
&=&U_t \exp\left( -\tau^2 \dot {E}^i_t \right)  \ . 
\ea
for small $\tau$, 
by definition of $P$ and the equation of motion for $U_t$. 
This completes the discussion of the Maxwell equations. 

For future reference, we rewrite eq.(\ref{flux}) in terms of the 
curvature tensor $F_{ij}$.

The smoothness assumption implies that the relation to continuum quantities
[temporarily distinguished by greek indices] is as follows:
\ba 
  A_i(x) &=&{\hat i}^{\mu }A_{\mu }(x) = O(a)  \\
      F_{ij}(x)&=&{\hat i}^{\mu }{\hat j}^{\nu}F_{\mu \nu }(x) = O(a^2) 
\ea
Smoothness of $A$ requires that a suitable gauge is chosen locally.

The discretization (\ref{U}) preserves the properties
\ba 
U(-b)&=& U(b)^{-1}  \\ 
U(\triangle ) &=& 
\exp\left(  - \frac 12 F_{ij}(x)+ ... \right) \label{2} \\
tr U(\triangle ) &=& 1+\frac 18 tr F_{ij}^2(x)+ ...  \label{3}
\ea
for a triangle $\triangle $ with corners $x, x+\hat i, x+\hat j$. To see this
one computes from the definition of the parallel transporters $U(\triangle ) = 
 1 - \frac 12\left( \partial_iA_j-\partial_jA_i + [A_i, A_j]\right)
(x)+O(a^3)$. It follows that eq.(\ref{2}) holds with dots representing 
terms which are of order $a^2$ and traceless. Eq.(\ref{3}) follows from this. 
All this is familiar from lattice gauge theory on a lattice of lattice spacing
$a$.   

The parallel transporters $U (b)$ represent the thin links $b$.
 In the case of general relativity, the vector potentials and the 
parallel transporters will be denoted by $\bfomega_i , \bf u$ in place of 
$A_i, U$.

\subsection{ General relativity in discrete time} 
\label{UnivGR}
 
In Ashtekar's canonical formulation of general relativity in the continuum,
the canonically conjugate variables (in the spinorial formulation) are
 as follows. 

There is a  vector potential  $\bfomega_i(x)\in sl(2,{\bf C})$ which governs 
parallel transport of spinors along paths on the space-like hyper-surface. 
The canonical conjugate to it is a spinorial version of the dreibein density, 
$2\tilde{\bf e}^i(x)\in sl(2,{\bf C})$. It is a traceless hermitian 
$2\times 2$ matrix. 
It is the analog of the electric field in Yang Mills theory.

The equations and motion are  
\ba 
2\dot {\bfomega}_k &=& 
 \underline N  [ {\bf F}_{ki},\tilde {\bf e}^i], 
 \label{EinsteinEqsCont1}  
 \\
2\dot  {\tilde {\bf e}}^k  &=&  
 \underline N D_j [\tilde {\bf e}^k, \tilde{\bf e}^j ] 
 \label{EinsteinEqsCont2}  \ea
The lapse function $ \underline{N}$ can be  chosen arbitrarily. $D_j$ is
the $sl(2,{\bf C})$-covariant derivative. 

The constraints are as follows. There is the Gauss constraint which 
is the same as in Yang Mills theory,
\be
D_i \tilde{\bf e}^i = 0 ,
\ee
In addition the following scalar and vector constraints hold
\ba
H &=&tr \left( {\bf F}_{ij}\tilde {\bf e}^i \tilde{\bf e}^j \right) = 0,
\\
H_j &=&  tr \left( {\bf F}_{ij}\tilde {\bf e}^i  \right) = 0,\\
\ea

The second equation of motion can be rewritten with the help of the Gauss
constraint as follows
\be
2\dot  {\tilde {\bf e}}^k  =  
 -\underline N  [ \tilde{\bf e}^j, D_j\tilde {\bf e}^k ].  
\ee
This brings about the dilemma which of the two versions to choose as 
a candidate precursor for a universal formulation. I choose 
eq.(\ref{EinsteinEqsCont2}) because it has a more natural geometric 
interpretation. If $\epsilon_{ijk}= \pm 1 $ as usual, depending on whether
$ijk$ is an even or odd permutation of $1,2,3$, and if 
one defines 
$$ f= \epsilon_{ijk}[{\bf \tilde e}^j, {\bf \tilde e}^k] dx^{i} $$ 
then the right hand side of eq.(\ref{EinsteinEqsCont2}) is
 expressible as a covariant total derivative of 
$f$. This parallels the situation in the Maxwell equations, where 
$curl B $ can also be regarded as total derivative of a 1-form 
$\epsilon_{ijk}B_{jk}dx^i$. 

A certain combination of the Gauss and vector constraint is supposed to ensure
the diffeomorphism invariance of the theory. It is a special property of
space, not of the universal dynamics which is supposed to be much more 
generally applicable. The equations which govern the universal dynamics have
 a   different kind of coordinate independence. 

There exists in the literature a lattice formulation of general relativity
in Ashtekar variables which is very similar to lattice gauge theory
\cite{loopRepres1}. 
In the case of the 2+1-dimensional theory there is a fully consistent 
discrete formulation of quantum gravity due to 
H. Waelbroeck \cite{Wael}. It uses the fact that the scalar and 
vector constraints in 2+1 dimensions are equivalent modulo invertibility
of the dreibein to 
$$ F_{ij}(x)=0 .  $$

In the lattice formulation, the vector potential gets replaced by 
parallel transporters
$u(b)\in SL(2,{\bf C})$. Apart from the gauge group this is 
 much as in the case of Yang Mills theory. The
parallel transporters  are assigned to links $b$ of the lattice.  

In the continuous time formulation, the dreibein on the lattice 
 remains an element of the Lie algebra. 
It is also assigned to links of the lattice (more precisely to 
one of
their endpoints; the dreibein  transform under gauge transformations like
 matter fields in the adjoint representation which sit on sites x).  
This is very similar to the electric field in the Hamiltonian formulation of 
lattice gauge theory \cite{Kogut}.

Let us again label the links $b=b_i(x)$ which enter a point $x$ of the lattice
by $i$. The link in opposite direction shall be $b_{-i}(x+\hat i) $.  
Henceforth, labels $i,j,k $ shall refer to links of the lattice. 
To avoid confusion, space indices will be labeled by $\mu , \nu, ...$ when
they are  needed. 
The basic variables are then $u(b_i(x))\equiv u_i(x)\in SL(2,{\bf C})$  and 
$\tilde {\bf e}^i(x)\in sl(2,{\bf C})$

A continuous time formulation needs much a priori structure; it seems 
impossible to formulate it without assuming at least that our spaces
 $\Omega_X$ are 
 manifolds. Because of this, time will also be discretized, and the 
dreibein will be exponentiated to
\be \E^i(x) =  \exp \left(-\tau {\bf e}^i(x)\right) \ee
$\tau $ will be related to the time step.
As a consequence of corresponding properties of the discretized dreibein
 variables \cite{Wael},
the $\E$-variables on adjoint links  are related by
\be
    u_i(x) \E^i(x) = \E^{-i}(x+\hat i) u_i(x). 
\ee
The Gauss law reads (to leading order in $\tau$ ) 
 $$ \prod_i \E^i(x)=1 $$
In the graphical representation the thin links represent parallel transporters
$u(b)=u_i(x)$ and the fat links are given by
\be  p(b)\equiv p_i(x)= u_i(x)\E^i(x) \label{p} \ee
They obey the unitarity relation  $p(b^{\ast})= p(b)^{-1}$.
The $\E$-variables get
 represented as hair pins, cp. the left hand
 side of figure \ref{GRGmotion} (2nd equation), and the 
Gauss  law takes again the form of 
 figure (\ref{Gauss}). 

In this approach the geometry is not in the lattice
 but in the parallel transporters on
the links of the lattice. For simplicity I will 
assume that charts  in the continuum which cover a sufficiently small
neighborhood of a point can be represented by a regular triangular mesh,
 as in figure \ref{grid}. In 3 space dimensions, a dense sphere packing 
can be used. It is assumed that the continuum vector potential is smooth 
on the scale of the mesh. This introduces a small parameter $\epsilon$.

To justify the discretized version of eq.(\ref{EinsteinEqsCont2}), 
the familiar relation between the group theoretical commutator 
$xyx^{-1}y^{-1}$ and the   Lie algebra commutator $[X,Y]$ is used. It yields
\be \E^j(x+\hat j)\E^k(x+\hat j)\E^j (x+\hat j)^{-1}\E^k(x+\hat j)^{-1} =
1 + \tau^2 [\tilde {\bf e}^j(x+\hat j),\tilde {\bf e}^{k}(x+\hat j)] \ee
The left hand side is represented by a composition of four hair-pins. They
can be seen in figure \ref{GRGmotion} (2nd half)
 except that the initial and terminal thin lines are missing.  
The links of the triangle in the figure are
$b_j(x), b_k(x+\hat j), b_{-i}(x+\hat i)$. 
The four hair-pins make a loop from $x+\hat j$ to $x+\hat j$. To be able to
compose these loops for different triangles with the same side 
$b_i(x)$, parallel transport from $x=\hat j$ to $x$ is needed. 
There are two paths (with one or two links) to choose from.
The difference  does not matter because its effect is of higher order in
 $\epsilon$. The  choice is made so as to minimize the total number of
 links in the path.

The rest of the argument goes as in the case of the Maxwell equation. 
The contribution from triangles on opposite sides of the link 
$b_i(x)$ will cancel out except for a contribution proportional
 to the covariant divergence. 

The other equation of motion and the constraints involve the curvature.
 There is an equation like eq.(\ref{magField}) 
which relates the parallel transporter
around a triangle to the magnetic field, except that now
 the analog of the magnetic field is the curvature
 of the $SL(2,{\bf C})$-connection on the space-like
hyper-surface. 

Consider 
\be u_{k,t+\tau} = u_{k,t} (1 -\tau \dot{\bfomega}_k + ...) \label{um}\ee
$u_k $ is represented by the horizontal thin lines in figure \ref{GRGmotion}.

The bitriangle in the first equation of figure \ref{GRGmotion} is 
\ba
\bitriangle_1  &=& u(\triangle)^{-1}u_j(x)^{\ast} p_j(x)u(\triangle)\\  
&=&  u(\triangle)^{-1}\left( 1 - \tau \tilde {\bf e}^j(x)+ ...\right)
 u(\triangle)\\
&=& 1 + \frac \tau 2 [\tilde {\bf e}^j (x), {\bf F}_{kj}] + ...  \\
&=& (1 -\tau \dot{\bfomega}_k+ ...)
\ea 
by eq.(\ref{2}) and Einstein equation (\ref{EinsteinEqsCont1}). 
Comparing, we see that the figure reproduces the time evolution (\ref{um}).

\section{The Dirac equation}
\label{Dirac}
The content of this section is based on joint work with 
B. Holm and D. L\"ubbert \cite{holmLubb}.

The massless Dirac or Weyl equation in flat 4-dimensional space time reads
\be -i\hbar \dot \psi = i\hbar c {\bfalpha^i }\cdot {\partial_i} \psi
\equiv \hbar c{\bf D}\psi \ee
with matrices $\alpha^i $ which obey the standard anti-commutation relations 
\be \{ \bfalpha^i, \bfalpha^j \}_+ = 2\delta^{ij}{\bf 1} \ (i,j=1...3)
\label{anticomm0}\ee
It follows from these relations that ${\bf D}$ is a square root of the 
negative Laplacian $-\Delta $. 

The appropriate inverse dreibein for a flat space is 
$ e_{a}^{\ i}= \delta_{a}^{\ i}$ with $ e^{-1}=det( e_{a}^{\ i})=1$.
It follows that the Ashtekar dreibein-variable in flat space  
\be {\bfalpha}^i = \tilde {\bf e}^i \equiv e  e_{a}^{\ j}\bfsigma^a \ee 
obeys the anti-commutation relations (\ref{anticomm0}). 
 
Let us denote by $-i$ the opposite direction to $i$;
 $\partial_{-i} =-\partial_{i}$.
  There is no distinct positive direction, and 
$\bfalpha^i\partial_i $ should have a meaning independent of choices of
positive directions. Therefore one should set
\be \bfalpha^i = -\bfalpha^{-i} \ . \ee 
The anti-commutation relations can now be written as 
\ba\label{anticomm}
\bfalpha^i\bfalpha^j &=& -\bfalpha^j\bfalpha^i \ \ (i\not=\pm j)\ ,  \\
\bfalpha^i\bfalpha^{-i} &=& -1 \ ,
\ea
and $\bfalpha^{2i}\equiv \bfalpha^i\bfalpha^i = 1$. We see here  -signs
which are of crucial importance, especially the
second one. The product is a product of matrices. 

In the Kogut Susskind discretization of the Dirac equation, 
 a cubic lattice with lattice sites $x$ 
 substitutes for continuous 3-space. In place of 
the Dirac $\bfalpha$-matrices one has 
numbers $\eta^i(x)$ attached to links $(x+\hat i, x)$ between  nearest 
neighbors $x$ and $x+\hat i$ on the lattice ($\hat i$ is the lattice 
vector in $i$-direction). In place of the Dirac algebra (\ref{anticomm}) one
has the relations 
\ba \label{eta}
\eta^i(x+j)\eta^j(x) & = & - \eta^j(x+i)\eta^i(x),\ (i\not=\pm j) \\
\eta^{-i}(x+\hat i) \eta^i(x)&=& -1 .  
\ea
Usually one requires in addition
\be \eta^{2i}(x)\equiv \eta^i(x+\hat i)\eta^i(x)=1 \label{unnec}. \ee 
But this relation can be abandoned. 
 The $\eta$-s sit on links. We may think of them as 
 parallel transporters of an (external) gauge field. (In the Kogut Susskind 
formalism the gauge group is ${\bf Z}_2$.)
Validity of eq(\ref{unnec}) can then be assured by a gauge transformation. 
This is seen as follows. 

 It follows from equations 
(\ref{eta}) that 
\be  
\eta^{2i}(x+2j)\eta^{2j}(x)  =   \eta^{2j}(x+2i)\eta^{2i}(x)
\label{2eta}
\ee
for all $i,j$. We 
may restrict attention to a sub-lattice of twice the lattice spacing, with
parallel transporters $\eta^{2i}$. Eq.(\ref{2eta}) tells us that $\eta^{2i}$
 is a pure gauge. Therefore it can be gauged away by a gauge transformation
$$\psi (x)\mapsto \gamma(x)\psi (x) ,\ \gamma (x)\in {\bf C}.$$ 
If the $\eta's$ are either numbers $\pm1$ or $\pm i$ then $\gamma (x) $ can 
be chosen in ${\bf Z}_2$. 

The Kogut Susskind discretization of the massless 
Dirac equation in continuous time is the $\delta t\mapsto 0 $ limit of  
the following equation
\be \psi (x, t+\delta t) = \psi (x,t) + \delta t
\sum_{i} \eta^{-i}(x+\hat i)\psi (x+\hat i) . \ee
The sum goes over all nearest neighbors  $x+\hat i$ of $x$, i.e. over
 positive and negative directions. $\hbar $ dropped out and we put $c=1$.
 The equation 
is invariant under the aforementioned gauge transformations. One may demand
$\eta^i(x)=\pm 1$. In this case the gauge freedom restricts to 
${\bf Z}(2)$-gauge transformations  $\gamma (x)= \pm 1$. The group ${\bf Z}(2)$
is the center  of $SL(2,{\bf C})$. It remains as a gauge group after  one
uses the flatness condition to gauge away the $SO(3,1)$-connection. 
 
 So far we considered a cubic lattice. The consideration can be generalized to 
a triangular lattice (dense sphere packing) \cite{holmLubb}. 
It turns out that  one needs to augment the
constraints (\ref{eta}) by the additional condition that the square of the 
parallel transporter around a triangle  is $(-1)$
in order to deduce the unfrustratedness condition \ref{2eta}. 
As a result the $\eta$'s
 take values $\pm i$. 

Now we wish to embed these formulae into the general frame work. 
In the case of the simple Dirac equation
in flat space and without gauge fields, the parallel transporter 
along the link 
from $x$ to $x+\hat i$  shall be given by the number $\eta^i(x)$ and $
\Omega_x$ is isomorphic to $\bf C$.  

As discussed before, we wish to think of the spinor fields $\psi (x)$ and
their adjoints $\bar\psi (x)$  as links to and from $\infty $. We parameterize 
these links by elements, also denoted   $\psi (x)$ of the linear space 
$\Omega_{x}$ and $\bar\psi (x)$ in its dual space, respectively.
We use the addition $+$ in these vector spaces in order to 
fix the composition law $\circinfty$ of arrows at $\infty $
 in a manner which will now be described. 

 We wish to 
exhibit properties of the initial state which guarantee that the universal
dynamics reduces to the Dirac equation. 
The composition law  is determined by the initial conditions.
Therefore we are free to select
an appropriate composition law, and to use other properties of the initial 
state to construct it. 

The composition law  $\circX $ of arrows at finite points $X$ of space is
 decreed to be 
given by the composition of maps between spaces $\Omega $, in accordance with 
the representation theorem. This theorem would tell us that there is also a 
space $\Omega_{\infty}$. But this is of no help, because this space could only 
be constructed once we know the composition law $\circinfty $. 
In accordance with the discussion in section (\ref{system}) 
 we construct this composition 
by imposing relations between paths.  

A general path which goes from $X$ to $Y$, meeting $\infty $ a number $N > 0$
 of times will have the form
$$ ... \psi \circinfty l_1\circinfty ... l_{N-1}\circinfty \bar\psi ... $$
where $\psi : \infty \mapsto X$, and  $\bar\psi (x): Y\mapsto \infty $ are 
links, and $l_i: \infty\mapsto \infty $ are paths.
The dots represent arrows which come from paths that do not touch $\infty$. 
 We impose the following relation. If $l_1 $ is a triangle of the form 
\be \psi^{\ast}\circ u_1 \circ \zeta_1 : \infty \mapsto \infty \label{trinf}\ee
where $u_1$ is a link between finite points, then  
\be 
\psi \circ l_1 =  \psi + u_1\circ \zeta_1   \ .  
\ee
while $\psi\circ \l_1 = 0 $ otherwise. 
If $l_1 ... l_M$ are of the form 
(\ref{trinf}) with different $u_i $ and $\zeta_i$ 
then the relation may be applied several times to produce a sum. 
$$\psi \circ l_1\circ ... \circ l_M =  \psi + \sum u_i\circ \zeta_i \  . $$  
Using this we see that the 
Kogut Susskind discretization of the 
Dirac equation can be written in the form of figure \ref{fundamentalEqMotion}.

If there is a gauge field, the $\Omega_x $ become 
representation spaces for the gauge group, and the $\eta^i $ 
get multiplied with parallel transporters 
that are furnished by the gauge field.
Only the first of the relations  (\ref{eta}) survives this, and the square 
of the covariant Dirac  operator is no longer equal to the negative Laplacian. 

 In curved space, the parallel 
transporters are associated with the Ashtekar variables $\omega_i $ and the 
$\eta^i$ involve the Ashtekar variables $ \tilde{\bf e}^i$. 
\section{Practical improvements} 
If one wants to use discretized versions of the Maxwell Yang Mills or Einstein 
Ashtekar equations in computer simulations, some improvement is called for. 
First, ons should {\em include growth processes in the dynamics} such that
the grid gets refined when the electric or magnetic field (or their analogue) 
get large. Secondly, one ought to guard against accumulating violations of the 
constraints from rounding or discretization errors by including suitable 
gradient terms which tend to restore the constraints. 

\section{Quantum Systems and Quantum Space Time}
\label{QSpaceTime}
Now we wish to proceed to a quantized theory. We use  Schroedinger  wave
functions. In standard quantum mechanics, they depend either on coordinates. 
or on momenta, but not both. Accordingly we assign
complex amplitudes 
\be \Psi (S)\in {\bf C} \ee
to systems $S$ which contain only one kind of links.
If we come from a classical description, we must choose 
 either thin links 
$b$ or fat links ${\bf b}$. In gravity we settle for the first possibility;
the thin links represent $SL(2{\bf C}) $ parallel transporters in this case. 

Before going into gravity, let us consider a simpler example.

{\bf Example } Quantum mechanical motion of particles in space\\
{\em
can be described by a universal dynamics.
We may picture a system of objects ("space points") linked by
bidirectional arrows plus additional objects ("point particles") linked
to space points (their positions) by one unidirectional arrow
 each, as in figure \ref{motion}.
The space is considered constant, the particles may move as indicated in 
figure \ref{motion}. In the case of a single particle , the system 
$S=S[x]$ is then determined by specifying the position $x$ of the particle and
 we may 
identify our Schr\"odinger amplitude $\Psi (S[x])$ with
a standard 1-particle Schr\"odinger wave function $\Psi (x)$.
 
A Schr\"odinger equation for the complex amplitude of such a system $S$
reads
\be i\dot{\Psi}(S)=
     -\sum_{\mu}[\Psi(\mu S)-\Psi(S)] \ .\ee
Summation
is over moves $\mu $
of individual particles as in figure \ref{motion}. For a single
particle of mass $m$
         on a triangular or cubic lattice of space points, this is the standard
discretization of the Schr\"odinger equation for free motion. To see this,
recall the standard discretization of the Laplace operator on a cubic lattice
of lattice spacing $a=1$,
$$ 
\Delta \Psi(x) = \sum_{\mu} [ \Psi (x + \hat{\mu})- \Psi (x)
] 
$$
where $x+\hat{\mu }$ is the nearest neighbour of $x$ on the lattice in $\mu $
 direction. Units of time $\hbar/2m$ are set to 1.}

At least in the case of a compact space, the dynamics of quantum gravity is
the most universal one that one can conceive of. There is no time dependence
because the Hamiltonian is zero. The interpretation of this fact is being
 debated. It is tempting to interpret Parmenides as saying that
 dynamics comes into
the world only through the separation between observer and observed. 
I come back to this, but for now let us accept that the only remaining 
problem in quantum gravity is to find the space of wave functions. 
And for any system is a nontrivial problem to find this space because
 of the constraints. 

The wave functions will be required to be gauge invariant. More generally, 
the wave
function cannot assign different amplitudes to systems $S$ which cannot be 
distinguished in an intrinsic way by properties which can be formulated in the 
language of thought. 

In the quantum theory,
gauge invariant functions of the parallel transporters will become 
multiplication operators.  Examples are operators $tr\, l$, where 
$l= b_1\circ ... \circ b_n$ are loops and 
where the trace $tr$ is a a gauge invariant function on loops. 
\footnote{
Such a function $tr$ always exists. An example is as follows.
 Consider loops $l: X\mapsto X$ and set
 $tr\, l = 1$ if $l=\iota_X$ and 0 otherwise. When $\Omega_X$ are 
linear spaces,  the trace of linear operators can be used instead. Remember 
that the existence of invariants like $tr$ is
 to be read off from the initial state}
  Operators which 
involve canonical momenta ("fat links") 
in the classical case will have to be represented as 
substitution operators.
They are called "grasp operators" in quantum gravity \cite{Ashtekar}. 
They map to linear combinations of wave functions 
for values of the argument $S$ which are obtained from each other by
some operation on $S$ which is specified by the observable. 

The interpretation of this prescription is somewhat subtle though.
 How would one specify 
a special loop $l$ in a general system $S$? In general this is impossible. 
But there are observables
like  $\prod tr \triangle $  (products over all triangular loops in $S$)
which are well defined products of such loops for all systems. 
Operators which have a physical meaning, such as the
 constraints, will have to be of this type. 

This gives us one immediate candidate for a solution to all
 conceivable constraints of the kind which demands
 invariance of the wave function under a particular possible dynamics.  
If the operation of the substitution operators on $S$ 
is compounded from local actions, they can not change the category of $S$.
[This generalizes the classical result of section \ref{conserv}.]
Therefore wave functions $\Psi (S)$ which depend on $S$ only 
through its category will obey all reasonable constraints.  
This reminds of the construction of thermodynamic ensembles from conserved 
quantities.

Let us specialize to quantum gravity. How can one label a basis in the space 
of wave functions $\Psi $? The problem has two parts. A system $S$ is specified
by a graph $G=${\em Graph}$(S)$ and  by relations between the paths on $G$.
In the traditional formulation, these relations are given implicitly by 
prescribing a gauge field configuration on $G$. The Mandelstam relations are
among these \cite{RovelliSmolin}.
The first problem then is to label wave functions $\psi_{G,p} $ whose
arguments are gauge field configurations, i.e. systems $S$ with a given graph
$G$. The second part of the problem concerns the generalization to wave 
functions which are defined for arbitrary systems $S$. I will restrict 
attention to wave functions whose support consists of systems whose graph
 is the 
skeleton of a simplicial complex. 

The first problem has recently been solved within the loop space approach to 
quantum gravity \cite{loopRepres1,loopRepres2} with the help of 
Penrose spin networks 
\cite{Penrose,RovelliSmolin,rigorous}. The issue had been to
 take account of the relations 
between traces of loops which hold true for every $SL(2,{\bf C})$ gauge field. 

A Penrose spin network $P$ is a graph with an assignment of a positive integer
$n(b)$, 
called its color,  to 
every undirected link $b$ of the graph, subject to certain conditions.
 If all the nodes
are trivalent, the condition is as follows. The sum of  the colors of 
the three links incident on a node has to be even, and none of them is larger
 than  the sum of the other two.  Since I prefer to work with directed graphs
where every link $b$ comes with a link $b^{\ast }$ in the opposite direction, 
I set 
$$ n(b)=n(b^{\ast}).$$
An embedding $p = \gamma_G(P)$ of a Penrose spin network in a graph $G$ is
a map of distinct nodes of $P$ to distinct nodes of $G$ and of links of $P$ 
between nodes
 to paths in $G$ between corresponding nodes. 
The paths must not intersect except as prescribed at the endpoints. 

To be in agreement with the literature, I change nomenclature. What has been 
called a loop up to now will henceforth be called a simple loop. And loops 
$\alpha $ are collections of simple loops ${ \alpha_1...\alpha_k}$. One sets
$$ tr \, \alpha = (-)^k\prod_{a=1}^ktr \, \alpha_a $$
To every embedded Penrose spin network $p$, a formal sum of loops $\alpha $ 
with coefficients $c_{\alpha }=\pm 1 $ is assigned by a certain prescription 
$$ p \Rightarrow 
\sum_{\alpha } c_{\alpha}\alpha $$
Let $l(b)$ be the number of times the path passes through link $b$ of $G$. 
The embedded spin network specifies an integer $n(b)\geq 0 $ for links of $G$. 
The prescription is such that $n(b) = l(b)+l(b^{\ast})$ and there is 
antisymmetry under operations of reconnecting the loops by
 permuting the end points of
instances of the same link $b$. For details, the reader is referred to the
 literature. The basis of states is given by
\be \Psi_{G,p} (S) = \sum c_{\alpha} tr \alpha \ee 
Given the graph $G$, one obtains in this way a linearly independent set
of gauge invariant functions of systems $S$ with given graph $G$. 

The action of loop operators on such states has been defined in
 the literature for loops with arbitrary number of
 "fat links" (dreibeins). In the above discussion
of classical gravity, an exponentiated dreibein was used, because we
had to mimic addition by use of the composition $\circ$. In quantum
gravity, there is no need for this, because the linearity of the state space 
supplies an operator $+$. The Lie algebra valued dreibein can be used.

We may adopt the construction, but there remains the 
above mentioned second problem to be 
faced. We do not wish to consider the space as preexisting (e.g. as a manifold)
with only a geometry that remains to to be put on it. Instead we want to
 define  wave functions $\Psi_p: S\mapsto {\bf C}$ on arbitrary systems $S$. 
There is no way to formulate an intrinsic specification
 of an embedded Penrose spin network which makes sense for a
 completely arbitrary system $S$. 
But there are distinguished classes of such embedded spin networks and we may 
sum or average $av$ over the representatives.  Let me write $av$ for this sum
or average.  If there is no representative
for a given system $S$, the result is zero. 

The solution of the diffeomorphism constraint in quantum gravity through the 
use of knots \cite{knots,loopRepres1} and its elaboration in terms of
 Penrose spin-networks \cite{RovelliSmolin} suggests what to do. 
The knot class $K$ of an 
embedded spin network is an equivalence class of spin networks 
which share an intrinsic property.  
It is called an $s$-knot. Embedded networks which are obtained from 
each other by homo\-topic deformation of the paths represent the same s-knot.

This leads us to try
\be \Psi_K(S)= N_K av_{p\in K}\Psi_{Graph(S),p}(S) \ee
where $N_K$ is an arbitrary normalization factor. 
These states are automatically diffeomorphism invariant. 

I wish to suggest a variation on this theme. Let us consider systems with a
distinguished object $O$, called its root. The idea is that all the states
are supposed to be subject to examination by one and the same observer
which is somewhere and thus marks a point in space. Briefly, call $O$
 the observer. Loops from $O$ to an object $X$ and back
 may be  interpreted as queries from $O$ to $X$ and answer back
 from $X$ to $O$. The message can be influenced
by the medium through which it passes. The construction above aims at
 labeling states by quantum numbers which record observable properties.
 If the observation is to be made by $O$, also the
 spin network which is to be embedded should
have a distinguished node, and this node is to be mapped on $O$.
This leads to "based knots". Parmenides and the idea of
considering a quantum system together with its observer suggest, moreover, 
to require gauge invariance only under gauge transformations which are 
trivial at $O$ and to admit functions $tr$ which are invariant in this
 restricted sense only and which may depend on the observer. This enlarges the 
space of observables. 

 Systems in
quantum mechanics which include an observer are not usually 
considered in text books. But they play a role in 
recent experiments on quantum erasers 
\cite{eraser}. These experiments show that  the effect of a measurement
 on a quantum system can be erased again, under certain conditions. 

\subsection*{Acknowledgment}
Part of this manuscript was written while the author visited the
 Hebrew University in Jerusalem and the Weizmann Institute of Science. 
I thank Sorin Solomon and Achi Brandt for their kind hospitality and for
 numerous discussions
on complex systems, and the 
German Israel Foundation for a travel grant. 
Discussions with Thomas Tomkos are also gratefully acknowledged. 

\section*{Appendix A: Proof of the main representation theorem}
A slightly more elaborate version of the representation theorem
was proven by the author in \cite{mac:1}. For the convenience of the
reader, the statement and proof of this theorem is reproduced here,
and it is shown how the  version in the main text, which is closer to
lattice gauge theory, is obtained as a corollary.
 
I use a slightly more elaborate notation than in the main text.
Given the system $K$
with objects $X,Y,...$, denote by  $\Mor (Y,X)$ the set of arrows
 $f:X\mapsto Y$ and
  $          g\circY f : X\mapsto Z \  $
for the composite of $f\in \Mor(Y,X)$ with $g\in \Mor(Z,Y)$.
  \begin{theo}\label{DarstSatz}{\em (Representation of a system as a
  communication network)} Every system  $K$ permits a faithful
  representation with the following properties
 
  To every object $X$ there exists an input space $\A_X $ and an output
  space $\O_X$. The input space contains a distinguished element
   $\emptyset$ ("empty input").
  \Arrow s $f\in \Mor (Y,X), g\in \Mor (Z,Y)$ and
  objects $X$ act as maps
  \ba X:\A_X&\mapsto& \O_X , \\
      \iota_X: \O_X &\mapsto & \A_X\\
      f:\O_X&\mapsto& \A_Y \ea
  with the properties
  \ba X\iota_X=id: \O_X \mapsto \O_X \ &,&
       \iota_X X =id: \A_X \mapsto \A_X \  ,\\
   g\circY f &=& gYf: \O_X \mapsto \A_Z \ . \label{CompMap}\ea
   \end{theo}
 It should be noted that  $\iota_X$ does not act as the identity map
in general in this context.

Given this version of the representation theorem, we restrict attention
to the output spaces $\O_X$ and to maps $\hat{f}= Y\circ f:
\O_X \mapsto \O_Y$. Renaming $\hat{f}$ into $f$
                    we obtain the representation theorem
of the main text.

\subsection*{Proof of the
                     representation theorem \ref{DarstSatz}
for categories}
Given a system $K$,
we write  $\Mor (Y,*)$ for the set of all its \arrow s
to $Y$ etc..
              We define
$$ In  (Y) = \Mor  (Y,*) \ \ , \ \ Out  (Y)= \Mor  (*,Y) \ . $$
We write $X=\a     (f)$ if  $f \in \Mor  (Y,X) \subset In  (Y)$,
and correspondingly
  $Z= \omega (f)$ if   $f\in \Mor  (Z,Y)\subset Out  (Y)$. The
output space will be defined as a subspace  $\Omega_Y$ of
                       $\Omega^{virt}_Y$.
$ \Omega^{virt}_Y$ consists of maps
$$ \zeta : Out_Y \mapsto \Mor  (*,*) $$
with the property  $\zeta (f) \in \Mor_K(\omega(f),*)$.
 
An object $Y$ will act as a map
$$ Y: In  (Y) \mapsto \Omega_Y . $$
according to
$$ Yf(g) = g \o Y f \ \ \ (g \in Out(Y)).  $$
The output space is defined as the image of $Y$, and the input space as
space of equivalence classes (if necessary)
                             of elements of $In_K(Y)$, which $Y$ maps
into the same  $\zeta \in \Omega_Y^{virt}$.
\ba  \Omega_Y&=& IM \  Y \subset \Omega^{virt}_Y \ , \\
     \A_Y &=& In  (Y)/ KER \  Y \ . \label{eqKA} \ea
$Y$ is invertible as a map from $A_Y$ to $\Omega_Y$. Its inverse
is  $\iota_Y$. The empty input $\emptyset \in \A_Y$ is defined as the
equivalence class of  $\iota_Y \in \Mor (Y,Y)\subset In (Y)$.
 
An \arrow \ $f \in \Mor(Y,X)$ is defined as a map  $\Omega_X \mapsto
\A_Y $ by use of the map $\iota_X: \Omega_Y
\mapsto \A_Y $, as follows.
\ba f &=& \hat f \o X  \iota_X \ , \\
    \hat f (g) &=& f\o X g \ \  \mbox{ for } g \in \Mor(X,*) \ .
\ea
The last formula defines  $\hat f $ as a map from $In (X)$ to
$In (Y)$. This map passes to equivalence classes (\ref{eqKA})
thereby defining a map                             $\A_X\mapsto A_Y$,
The composition rule
 (\ref{CompMap}) holds.
\section*{Appendix B: General Relativity in Ashtekar variables} 
For the convenience of the reader I will briefly review the 
canonical formalism for general relativity in Ashtekar variables.

Before dealing with Ashtekar variables, let me briefly examine classical
general relativity in order to see what a priori structural assumptions
are made.

In classical general relativity
one deals with a 4-dimensional space time manifold $\M$ 
and with 
a dynamically determined geometry on $\M$.
 The geometry  provides a 
connection in the tangent bundle which is compatible with a Lorentzian metric.
Field equations for the metric and the connection are derived from 
a variational principle. 
The vanishing of the torsion is one of these field equations.

In the vierbein formalism,  the connection in the 
tangent bundle can be thought to be constructed in two steps.
\begin{enumerate}
\item 
 There is a
 connection in a vector bundle over $\M $ with fibers
 $V_x\approx V^{(\half , \half)}\approx {\bf R}^4 $. 
This connection preserves a bilinear form $ < , >_x $ of signature 
$(+---)$ on the fibers.
\item
The fibers $V_x$ are identified with the tangent spaces $T_x\M$.
\end{enumerate}
$V^{(\half , \half)}$ is the 4-dimensional real representation space of 
the Lorentz group (``vector representation'').

The connection specifies parallel transporters. They are linear maps
which preserve the bilinear form
\ba  \UC &:& V_x \mapsto V_y, \\
 <\UC v , \UC w >_y &=& <v , w>_x 
 \ea
They transport  vectors $v, w\in V_x $ along piecewise smooth paths $C$ on
from $x$ to $y$. 

The identification is provided by a vierbein. It specifies an 
invertible map from the tangent space to the internal space
\footnote{Beware of confusion. This is not the same $\E $ as in the main text.}
\be {\cal E}(x): T_x\M \mapsto V_x \ee 
 for every $x$.
By virtue of the identification, the bilinear form $<, >_x$ on the fibers
becomes a Lorentz metric $g$  on $\M $, viz.
 $g(X,Y)=<{\cal E}(x)X,{\cal E}(y)Y>_x$
for $X,Y\in T_x\M $.

In this manner, general relativity appears as a gauge theory with gauge group
isomorphic to the Lorentz group $SO(3,1)$ and with a distinguished field, the 
vierbein field. The action has a particular form.

The standard description is obtained by introducing  coordinate systems on 
charts of 
$\M $ and a moving frame on each chart. The moving frame specifies 
a pseudo-orthonormal basis ${\bf f }(x)= (f_0(x),f_1(x),f_2(x),f_3(x))$ 
of $V_x$ for every $x$ in the chart. Pseudo-orthogonality reads
$$ <f_{\alpha}(x), f_{\beta}(x)>_x  = \eta_{\alpha \beta} $$
with $\eta_{\alpha \beta} =diag(+1,-1,-1,-1)$. The moving frame serves to
convert linear maps into matrices.  

The pseudo-orthonormal frames ${\bf f}(x)$ form the fibers of a principal 
fiber bundle whose structure group is the Lorentz group $SO(3,1)$. Parallel
transport of vectors induces parallel transport of frames and thereby a
connection on a principal fiber bundle.
  
The coordinate system specifies a basis $\partial_{\mu} $ in the tangent spaces
$T_x\M$. Expanding everything in sight,
 one gets the components of the vierbein and of its inverse , the
 components of the metric tensor,
 and the parallel transport matrices 
${\bf U} (C) \in SO(3,1)$ with entries $U^{\alpha}_{\ \beta}$. 
\ba
{\cal E}(x)\partial_{\mu } & = & E_{\mu}^{\ \alpha }(x)f_{\alpha }(x) ,\\
 {\cal E}(x)^{-1}f_{\alpha }(x)&=& 
  E_{ \alpha }^{\ \mu }(x)\partial_{\mu },\\  
E_{\mu }^{\ \alpha} \eta_{\alpha \beta} E_{\nu}^{\ \beta} &=& g_{\mu \nu }(x) 
\\ 
 \UC f_{\alpha }(x) &=& f_{\beta}(y)U^{\beta}_{\ \alpha}(C) . 
\ea 
The moving frame serves to convert linear maps into matrices. 
The parallel transport matrix $\bfU (C) $ for infinitesimal paths $C$ from
a point $x$ with coordinates $x^{\mu }$ to a neighboring point with 
coordinates $x^{\mu }+\delta x^{\mu }$ defines the vector potential
$\Gamma_{ \mu}(x)= (  \Gamma^{\alpha}_{\ \beta \mu}(x)  )$
\be {\bf U}(C)= 1 - \Gamma_{\mu}(x)\delta x^{\mu } . \ee
The entries of the vector potential are also known as
 the connection coefficients  in the anholonomic
basis provided by $f_{\alpha }$. 

The matrix ${\bf R}_{\mu \nu }$ whose entries are the 
 anholonomic components $R^{\alpha}_{\ \beta \mu \nu } $ of the 
field strength- or curvature-tensor $\F_{\mu \nu }(x): V_x\mapsto V_y $ 
are given by the standard formula
\ba 
\F_{\mu\nu }(x) f_{\alpha}(x)&=& f_{\beta }(x)R^{\beta}_{\ \alpha \mu \nu }(x)
, \\
{\bf R}_{\mu \nu }&=& \partial_{\mu }\Gamma_{\nu } -
\partial_{\nu }\Gamma_{\mu } +\Gamma_{\mu }\Gamma_{\nu } -
\Gamma_{\nu }\Gamma_{\mu }.  
\ea
One may compute the parallel transporter $\UC   $ around an infinitesimal 
triangle $C=\bigtriangleup $ whose corners have coordinates 
$\{ x^{\mu }\} , \{ x^{\mu }+ \delta  y^{\mu } \} , 
 \{ x^{\mu }+ \delta  z^{\mu } \}. $ 
 The result can be stated in basis independent form as 
\be \U (\bigtriangleup ) = 1 -\half \F_{\mu \nu }(x) \delta  y^{\mu } 
 \delta  z^{\nu } + ... \ .
\ee
The differentials $ \delta  y^{\mu }, \delta  z^{\nu }$ should be regarded as 
anti-commuting. 

Gauge transformations are determines by matrices
 $S(x)= (S^{\alpha}_{\ \beta}(x)) \in SO(3,1)$. A (passive) gauge
 transformation is a change of moving frame
\be f_{\alpha}(x) \mapsto f_{\beta}(x)S^{\beta}_{\ \alpha}(x). \ee
This transformation preserves pseudo-orthonormality. 
The parallel transport matrix,
vierbein components and vector potential transform in the familiar way under
 such gauge transformations.
\subsection*{Connections in spinor space. Ashtekar variables}
The Ashtekar variables appear very naturally if one starts   from 
parallel transport of spinors rather than 4-vectors. Such parallel transport
of spinors must be considered anyway when one wants to describe matter
by wave functions for spin     $\half$ particles. The gauge group 
is then the quantum mechanical Lorentz group $SL(2,{\bf C})$.  

Because of the structural assumptions of the standard theory,  
the parallel transport of vectors in 
an arbitrary representation space of the structure
group determines the parallel transport of vectors in any representation 
space.

The fibers  $V^+_x \approx V^{(\half , 0)} \approx \C^2 $ and 
 $V^-_x \approx V^{(0, \half )}  \approx \C^2 $ are now isomorphic to 
2-dimensional complex representation spaces of $SL(2,{\bf C})$. One has
\be
V^{(\half , \half })
 = V^{(\half , 0)} \otimes V^{(0 , \half )} . \label{tens}
\ee
More precisely, $V^{(\half , \half })$ is a real subspace of the complex
representation space $ V^{(\half , 0)} \otimes V^{(0 , \half )}$.
This identification can serve to construct a moving frame in 
$V^{(\half , 0)} \otimes V^{(0 , \half )}$ from a moving frame in 
$V^{(\half , \half })$. The parallel transport matrices 
${\bf u}(C)\in SL(2,{\bf C})$ 
 for vectors in $V_x^+\approx V^{(\half , 0)} $ and 
$U (C)\in SO(3,1) $  for vectors in $V_x\approx V^{(\half , \half )} $ 
are related 
by the fundamental formula of spinor calculus,
\ba
A {\bfsigma}_{\alpha }A^{\ast } &=&
{\bfsigma}_{\beta }\Lambda^{\beta}_{\ \alpha}(A)
\ \ \ \mbox{for } {\bf A}\in SL(2,{\bf C}) , \label{fundspin} \\
{\bf X} {\bfsigma}_{\alpha } +{\bfsigma}_{\alpha }{\bf X}^{\ast }
 &=& {\bfsigma}_{\beta }\Lambda^{\beta}_{\ \alpha}({\bf X} )
\ \ \ \mbox{for } {\bf X}\in sl(2,{\bf C}) .
\label{fundspinLie}  
\ea ${\bfsigma}_i$ are  the Pauli 
matrices for  $i=1,2,3$ and ${\bfsigma}_0 = {\bf 1} $
 ($2\times 2$ identity matrix). This formula yields the Lorentz transformation 
$\Lambda({\bf A})$ which is associated with ${\bf A} \in 
SL(2,{\bf C})$, and similarly for elements of the Lie algebra $sl(2,{\bf C})$. 

We will use boldface letters to characterize complex $2\times 2$ matrices 
throughout. 

The parallel transport matrices  ${\bf u}(C)$, the 
vector potential ${\bfomega}_{\mu }\in sl(2, {\bf C}) $ and the field
 strength matrix
 ${\bf F}_{\mu \nu}\in sl(2,{\bf C})$  in spinor space obey the following 
relations
\ba
{\bf u}(C) \bfsigma_{\alpha }{\bf u}(C)^{\ast }
 &=& \bfsigma_{\beta }U(C)^{\beta}_{\ \alpha}, \\
{\bf F}_{\mu \nu }(x)
 \bfsigma_{\alpha } +\bfsigma_{\alpha }{\bf F}_{\mu \nu }(x)^{\ast}
 &=& \bfsigma_{\beta }R^{\beta}_{\ \alpha \mu \nu }(x )\label{RF}\\
{\bf u}(C) &=& 1 - {\bfomega}_{\mu }(x) \delta x^{\mu } + ... \\
{\bf F}_{\mu \nu } &=& \pl_{\mu }{\bfomega}_{\nu } -  
  \pl_{\nu }{\bfomega}_{\mu }
+{\bfomega}_{\mu }{\bfomega}_{\nu }-{\bfomega}_{\nu }{\bfomega}_{\mu }. 
\ea 
One converts also the vierbein ${\cal E}$ to a spinorial basis. For a fixed 
$\mu $, the vector ${\cal E}(x)\pl_{\mu } \in V_x $. Because of the
 identification (\ref{tens}) of representation spaces, we may also regard
${\cal E}(x)\pl_{\mu }$ as an element of $V^+_x\otimes V^{-}_x $. 

To obtain convenient formulas, one 
 introduces  ${\tildebfsigma }_i = - {\bfsigma}_i\ (i=1,2,3) $ and 
${\tildebfsigma}_0 = {\bfsigma}_0 $ so that
\ba
{\bfsigma}_{\mu}{\tildebfsigma}_{\nu} +
 {\bfsigma}_{\nu}{\tildebfsigma}_{\mu}& = &
2\eta_{\mu \nu }, \\
\epsilon \ ^t{\bfsigma}_{\mu } \epsilon^{-1} &=& \tildebfsigma_{\mu } . 
\ea
where $\epsilon $ is the completely antisymmetric tensor in two dimensions, 
and $\ ^t\bfsigma $ is the transpose of the matrix  $\bfsigma $. 

We introduce two (hermitian) $2\times 2$-matrices ${\bf E}^{\mu }$ and
$$\tilde{\bf E}^{\mu }= \epsilon \ ^t{\bf E}^{\mu }\epsilon^{-1}$$ 
 as spinorial versions of the inverse vierbein
$(E_{\alpha }^{\ \mu })$.
\ba
E_{\alpha }^{\ \mu }&=& tr \left(  \bfsigma_{\alpha }
 \tilde{\bf E}^{\mu }\right)=
 tr \left(\tildebfsigma_{\alpha } {\bf E}^{\mu }\right), \\
\bfsigma_{\beta } E^{\beta \mu } &=& {\bf E}^{\mu } . 
\ea
The Einstein action will involve the  volume form $E(x)d^4x$
with $$E= det(E_{\mu}^{\ \alpha}) = det (E^{\ \mu}_{\alpha})^{-1} = 
\sqrt (-g).$$ Using the relation (\ref{RF}) of the field strength and the 
definitions of the spinorial versions of the vierbein one finds
\ba \mbox{\rm Einstein action} &=& \int d^4x L ; \\
L &=& E E_{\alpha }^{\ \mu }E^{\beta \nu }  
R^{\a}_{\ \b \mu \nu }\\
& = & E\ tr \left\{ 
  {\bf F}_{\mu \nu}{\bf E}^{\nu }\tilde {\bf E}^{\mu } + {\bf E}^{\ast \nu }  
  {\bf F}_{\mu \nu}^{\ast } \tilde {\bf E}^{\ast \mu } \right\} . 
\label{Einstein}
\ea
By a choice of gauge, i.e. of a suitable moving frame, the inverse 
vierbein is brought to the form
\be 
\left(
\begin{array}{cc}
N^{-1}  & -N^{-1}N^m \\
0 & { e}_{a}^{\ i} 
\end{array} 
\right)
\ee 
It follows that
\be
E=Ne , \ \ \ e=det(e_{a}^{\  i })^{-1} . 
\ee
$N$ and $N^{m}$ are known as lapse and shift functions, and 
$({\bf e}_{a}^{\ i})$ is the  inverse dreibein,
with holonomic indices $i=1,2,3$ and anholonomic indices $a=1,2,3$.  
Its spinorial version ${\bf e }^i$ is given by 
\ba
 e_{a}^{\ i} &=& \half tr \left( \tildebfsigma_{a }{\bf e}^{i}\right),\\
{\bf E}^i &=& {\bf e}^i + E_0^{\ i}{\bfsigma}_0 . \label{edec}
\ea
Finally one introduces
\ba
\underline N &=&  Ne^{-1}  \\
\tilde {\bf e}^i &=& e {\bf e}^i . 
\ea
Consider now a space-like surface $x^0 = t$. The parallel transport  
along curves $C$ within the surface is determined by the space components 
of $\bfomega_i $ of the vector potential. 

The Ashtekar variables are the $2\times2$ matrices $\bfomega_i$ and 
$\tilde {\bf e}^i $.  
\subsection*{Canonical formalism}
Dirac's canonical formalism with constraints is used to bring the equation of 
motion to Hamiltonian form and to determine  constraints on the initial
data. There is a subtle point, however. Ashtekar 
applies the canonical formalism to a theory 
with a gauge group $SL(2,{\bf C})\times SL(2,{\bf C})$ ("complex relativity").
This means that left handed and right handed spinors have {\em independent}
 parallel transporters, with independent vector potentials $\bfomega $ and
$\bfomega^{\ast}$, and there are also two independent vierbein variables
$E$ and $E^{\ast}$. 
This is quite natural from the point of view of the philosophy of this paper. 
To compare the left and right handed spinors, one would need 
to assume that there exists an invariant operation, complex conjugation, 
which commutes with parallel transport. It is natural to consider 
the existence of such an invariant and 
an associated relation between parallel transporters as a
 distinguishing feature of initial states in general relativity, but not 
as {\em a priori} structure. 

Dirac's formalism is described in detail in the text book \cite{Teitelboim}. 
 In the course of the analysis, first class and second class constraints 
appear. The second class constraints must be imposed as strong constraints,
i.e. they are equalities on all the phase space.
 
As a warm up exercise the reader may consider the following nonstandard
 form of the Maxwell action 
\be 
L_{Maxwell} = -  \int d^4x
 \left\{ (\partial_{\mu} A_{\nu } -\partial_{\nu} A_{\mu })F^{\mu \nu}   
-\half F_{\mu \nu}F^{\mu \nu} \right\} 
\ee
The condition $F_{ij}= \partial_{i} A_{j }    -  \partial_{j} A_{i }$ 
arises in the form of a second class constraint.
 The standard Hamiltonian is obtained
and the Gauss law is a first class constraint. 

Let us return to the Einstein action.
${\bfomega}_{\mu }$ are complex traceless $2\times 2$ matrices. Let me repeat: 
Ashtekar deviates from the rules in the book in an important way. 
He starts with the assumption that the $sl(2,{\bf C})$ vector potentials 
$\bfomega $ and ${\bfomega}^{\ast }$ are independent variables to begin with, 
and also the dreibein variables which come to multiply 
${\bf F}_{\mu \nu}$ and ${\bf F}_{\mu \nu}^{\ast}$ in the formula for the 
Einstein action are independent variables. In other words, he starts with 
complex general relativity; the reality constraints are only imposed 
to select among the solutions. 

Proceeding in this way, attention is restricted to the first term in the 
Einstein action (\ref{Einstein}) which involves $\bfomega_{\mu}$.  
The first step is the determination of the conjugate variables
$\bfpi^{\mu }$ .  One finds
\be
 \ ^t\bfpi^{\mu } = \frac {\pl L}{\pl \pl_0{\bfomega}_{\mu }}=
 E\left( {\bf E}^{\mu }\tilde{\bf E}^0 - {\bf E}^0\tilde {\bf E}^{\mu }\right) 
\ee
$\ ^t\bfsigma $ stands for the transpose of a $2\times 2$ matrix $\bfsigma $
 again. 

One deduces from this that
 $\bfpi^0 = 0 $. In addition one finds
Ashtekar's celebrated result that vector potential and the
``densitized'' inverse dreibein are conjugate variables
\be
\bfpi^i = 2\tilde {\bf e}^i . \label{can} 
\ee 
Proceeding further according to the rules in the book, the 
canonical Hamiltonian $H_C$ is obtained as a sum of three terms, all of 
which are first class constraints. This means that they vanish on the 
constraint surface. 
\ba
H_C &=& -\underline N H - 2N^i H_i - 2 G , \\
H &=&tr \left( {\bf F}_{ij}\tilde {\bf e}^i \tilde{\bf e}^j \right),
\\
H_j &=&  tr \left( {\bf F}_{ij}\tilde {\bf e}^i  \right),\\
G &=& tr \left( {\bfomega}_0 D_i \tilde{\bf e}^i \right) \ . 
\ea
Summation are from 1 to 3. The hermitian conjugate $h.c.$ involves the 
adjoint matrices ${\bf F}_{\mu \nu}^{\ast}$ and ${\bfomega}_0^{\ast }$; 
the dreibein matrices $\tilde {\bf e}^i$ are self adjoint. 

This is the second one of Ashtekar's celebrated results. It exhibits 
the constraints in polynomial form. $\underline N$, $N^i $ and the matrix 
$\bfomega_0$ are Lagrange multipliers; they have zero conjugate variables. 
The covariant derivative has the conventional form
\be D_i {\bfsigma} = \pl_i{\bfsigma} + [{\bfomega}_i ,{\bfsigma}] . \ee
Time development  is from space-like hyper-surface to space-like 
hyper-surface. 
There is freedom in choosing the foliation into hyper-surfaces
and of Lorentz gauge transformations. This is  reflected 
in the freedom of choice of the Lagrange multipliers. We choose
\be  \ \ N^i = 0, \ \ \ {\bfomega }_0= 0 . \ee
The field equations are obtained with the help of the following lemma. 
Suppose ${\bf A}^{ij}$ is independent of $\bfomega_k$. Then
\be 
\frac {\delta}{\delta {\bfomega}_k} tr \left({\bf A}^{ij}{\bf F}_{ij}\right)   
=\ ^t\left( D_j{\bf A}^{kj}-D_j{\bf A}^{jk}\right) . 
\ee
One obtains
\ba 
2\dot {\bfomega}_k &=& \frac {\partial H_C}{\partial \tilde{\bf e}^k}
= \underline N  [ {\bf F}_{ki},\tilde {\bf e}^i], 
 \\
2\dot  {\tilde {\bf e}}^k  &=&  
  - \frac {\partial H_C}{\partial {\bfomega}_k}
= \underline N D_j [\tilde {\bf e}^k, \tilde{\bf e}^j ] 
\label{EinsteinEqsCont}  \ea

\end{document}